\documentclass[fleqn,usenatbib]{mnras}

\usepackage{newtxtext,newtxmath}
\usepackage{bm}
\usepackage{float}
\usepackage{stfloats}
\usepackage{graphics}
\usepackage{appendix}
\usepackage{subcaption}
\usepackage{multirow}
\usepackage{graphicx}    % Include figure files
\usepackage{dcolumn}     % Align table columns on decimal point
\usepackage{bm}          % bold math
\usepackage{color}
\usepackage{tikz}
\usetikzlibrary{positioning}
\usepackage[T1]{fontenc}
\usepackage{xspace}
\usepackage{etoolbox}
\usepackage{tikz,xcolor,hyperref}% Make Orcid icon
\usepackage{graphicx}	% Including figure files
\usepackage{amsmath}	% Advanced maths commands
\usepackage{enumitem}
\usepackage{soul}

\DeclareRobustCommand{\VAN}[3]{#2}
\let\VANthebibliography\thebibliography
\def\thebibliography{\DeclareRobustCommand{\VAN}[3]{##3}\VANthebibliography}

\newcommand{\hi}{\text{H\textsc{i}}\xspace}

\newcommand{\secref}[1]{\hyperref[#1]{Section~\ref*{#1}}}
\newcommand{\appref}[1]{\hyperref[#1]{Appendix~\ref*{#1}}}

\preto\align{\par\nobreak\noindent}
\expandafter\preto\csname align*\endcsname{\par\nobreak\noindent}
\preto\multline{\par\nobreak\noindent}
\expandafter\preto\csname align*\endcsname{\par\nobreak\noindent}
\preto\flalign{\par\nobreak\noindent}
\expandafter\preto\csname align*\endcsname{\par\nobreak\noindent}
\preto\eqnarray{\par\nobreak\noindent}
\expandafter\preto\csname align*\endcsname{\par\nobreak\noindent}

\newcommand*{\msk}{\\[0.25cm]} % mathlineskip
\newcommand*{\nmsk}{\notag\msk} % mathlineskip with no label
\definecolor{lime}{HTML}{A6CE39}
\DeclareRobustCommand{\orcidicon}{%
    \begin{tikzpicture}
    \draw[lime, fill=lime] (0,0) 
    circle [radius=0.16] 
    node[white] {{\fontfamily{qag}\selectfont \tiny ID}};    \draw[white, fill=white] (-0.0625,0.095) 
    circle [radius=0.007];    \end{tikzpicture}
    \hspace{-2mm}}
\foreach \x in {A, ..., Z}{%
    \expandafter\xdef\csname orcid\x\endcsname{\noexpand\href{https://orcid.org/\csname orcidauthor\x\endcsname}{\noexpand\orcidicon}}
    }
\title[\hi\ delay spectrum with MeerKAT interferometer mode]{\hi\ intensity mapping with MeerKAT: forecast for delay power spectrum measurement using interferometer mode}
\author[M. Zhang et al.]{Ming Zhang\orcidA{}$^{1}$, Yichao Li\orcidB{}$^{1}$\thanks{E-mail: liyichao@mail.neu.edu.cn}, Jing-Fei Zhang\orcidC{}$^{1}$, Xin Zhang\orcidD{}$^{1,2,3}$\thanks{E-mail: zhangxin@mail.neu.edu.cn}
\\
$^{1}$Key Laboratory of Cosmology and Astrophysics (Liaoning) \& College of Sciences, Northeastern University, Shenyang 110819, China\\
$^{2}$Key Laboratory of Data Analytics and Optimization for Smart Industry (Ministry of Education),
Northeastern University, Shenyang 110819, China\\
$^{3}$National Frontiers Science Center for Industrial Intelligence and Systems Optimization,
Northeastern University, Shenyang 110819, China\\
}

\pubyear{2023}

\begin{document}
\label{firstpage}
\pagerange{\pageref{firstpage}--\pageref{lastpage}}
\maketitle

% Abstract of the paper
\begin{abstract}
Neutral hydrogen (\hi) intensity mapping (IM) surveys are considered a promising tool for investigating the expansion history of the Universe. In this work, we explore the potential of MeerKAT \hi IM observations in interferometer mode to estimate the power spectrum and constrain cosmological parameters within typical dark energy models. We employ an approach called the ``delay spectrum,'' which allows us to separate the weak \hi signal from foreground contamination in the frequency domain. 
Our findings indicate that the choice of survey fields significantly impacts the fractional errors on the power spectrum ($\Delta P/P$) within a limited observational time of 10 hours. As the integration time increases from 10 hours to 10,000 hours, $\Delta P/P$ progressively decreases until cosmic variance begins to dominate. For a total observation time of 10,000 hours, the lowest $\Delta P/P$ at low $k$ can be achieved by tracking 100 points for MeerKAT L-band (900--1200 MHz) and 10 points for MeerKAT UHF-band (580--1000 MHz).
Next, we assess the performance of \hi IM in constraining typical dark energy models. We find that MeerKAT \hi IM survey in interferometer mode demonstrates limited capability in constraining the dark-energy equation of state, even when combined with \emph{Planck} data.
Our analysis serves as a valuable guide for future MeerKAT observations in \hi IM surveys.

\end{abstract}

% Select between one and six entries from the list of approved keywords.
% Don't make up new ones.

%%%%%%%%%%%%%%%%%%%%%%%%%%%%%%%%%%%%%%%%%%%%%%%%%%

%%%%%%%%%%%%%%%%% BODY OF PAPER %%%%%%%%%%%%%%%%%%

\section{Introduction}
In the last two decades, the accurate measurement of the cosmic microwave background (CMB) brings 
us to the era of precision cosmology. Another promising method for reaching precision cosmology 
is the cosmic large-scale structure (LSS) survey. At present, the cosmic LSS survey has made significant 
progress with galaxy redshift survey in constraining cosmological parameters, e.g. 
the 2dF Galaxy Redshift Survey \citep{2DFGRS:2001zay,2dFGRS:2005yhx}, 
the 6dF Galaxy Survey \citep{Jones:2009yz,Beutler:2011hx}, 
the WiggleZ Dark Energy Survey \citep{Blake:2011en,Drinkwater:2009sd}, 
the Baryon Oscillation Spectroscopic Survey (BOSS) \citep{BOSS:2016wmc} 
and the Dark Energy Survey (DES) \citep{DES:2017myr}. 
In addition, the next generation galaxy survey targeting an even larger and deeper Universe,
such as the Dark Energy Spectroscopic Instrument \citep{DESI:2018ymu}, 
the Large Synoptic Survey Telescope \citep{LSST:2008ijt,LSSTDarkEnergyScience:2018yem}
and the \emph{Euclid} \citep{Amendola:2016saw}, 
will significantly improve the measurement precision in the near future.

Neutral hydrogen (\hi) is widely regarded as a promising tracer of the underlying dark matter 
distribution of the late Universe. \hi can be detected with radio telescopes via its 21 cm line 
which arises from the spin-flip transition of ground state hydrogen atom. 
Nevertheless, it is known that detecting \hi signal in individual galaxies at higher redshift 
requires good angular resolution and sensitivity, which relies on 
large radio interferometers in the near future, such as the Square Kilometre Array (SKA). 
However, \hi\ survey of the cosmic LSS can be quickly carried out using existing radio telescopes 
via the intensity mapping (IM) methodology, which observes the total \hi intensity of the galaxies in a voxel
\citep{Battye:2004re,McQuinn:2005hk,Loeb:2008hg,Chang:2007xk,Wyithe:2007rq,Bagla:2009jy,Seo:2009fq,Lidz:2011dx,Ansari:2011bv}. 
A variety of related studies show that \hi IM survey has great potential for cosmology studies, 
e.g. constraints on cosmological parameters and related studies on dark energy
\citep{Pritchard:2011xb,Bull:2014rha,Pourtsidou:2016dzn,Olivari:2017bfv,Obuljen:2017jiy,Sprenger:2018tdb,Xu:2017rfo,Cheng:2019bkh,Xu:2020uws,Jin:2020hmc,Xiao:2021nmk,Zhang:2021yof,Jin:2021pcv,Wu:2021vfz,Scelfo:2021fqe,Berti:2021ccw,Wu:2023wpj,Wu:2022jkf}, 
primordial non-Gaussianity \citep{Camera:2013kpa,Xu:2014bya,Li:2017jnt,Ballardini:2019wxj,Karagiannis:2019jjx,Cunnington:2020wdu,Karagiannis:2020dpq,Viljoen:2021ypp}, and neutrino mass \citep{Villaescusa-Navarro:2015cca,Zhang:2019ipd}.

\hi IM LSS detection was first reported by measuring the cross-correlation function between the 
\hi\ map observed with Green Bank Telescope (GBT) and the DEEP2 optical redshift survey \citep{Chang:2010jp}. 
The cross-correlation power spectrum between \hi IM survey and optical galaxy survey was also detected by 
the GBT and the WiggleZ Dark Energy Survey \citep{Masui:2012zc}. 
\citet{Anderson:2017ert} reported the cross-correlation power spectrum result of the Parkes Telescope's \hi IM map and the 2dF galaxy survey. With the same dataset, \citet{Tramonte:2020csa} presented the 
feasibility of measuring \hi filament with the intensity mapping survey.
%A cross-correlation signal detected with Parkes \hi\ IM and WiggleZ redshift data was discussed in \citet{Li:2020pre}. 
\citet{Wolz:2021ofa} gave a joint analysis of GBT \hi IM and eBOSS survey. 
Recently, the MeerKAT \hi IM survey 
reported the cross-correlation power spectrum detection of the \hi map and the
optical galaxy survey \citep{2022arXiv220601579C}. 
Meanwhile, using the MeerKAT interferometer observations, 
\citet{2023arXiv230111943P} reported the \hi IM auto power spectrum detection 
on Mpc scales. The \hi IM auto power spectrum on large scales, however, still 
remains undetected \citep{2013MNRAS.434L..46S}. 
%To date, the auto-correlation power spectrum is still not detected because of systematics and foreground contamination. 
%Most recently, it was also proposed to have an \hi IM survey with the newly built
%MeerKAT telescope in single-dish mode \citep{MeerKLASS:2017vgf,2021MNRAS.501.4344L,Wang:2020lkn}
%reported the cross-correlation power spectrum detection with the
%optical galaxy survey \citep{2022arXiv220601579C}. 

Several large radio telescopes or interferometers, such as the Five-hundred-meter Aperture Spherical radio Telescope 
\citep[FAST,][]{Nan:2011um}, Baryon acoustic oscillations In Neutral Gas Observations 
\citep[BINGO,][]{Dickinson:2014wda,Wuensche:2019znm} and SKA \citep{Maartens:2015mra,SKA:2018ckk}, 
are built or planned for \hi\ survey.
%which are expected to detect the auto-correlation power spectrum.
%However, interferometer mode (cross-correlation), compared to single-dish mode, has inherent advantages. Besides providing high angular resolution, interferometers are less sensitive to systematics which is a major problem to the auto-correlation power. 
In addition, interferometers have the inherent advantage of being less sensitive to systematics.
%which can be regarded as a complementary approach to single-dish IM experiments.
The smallest $k$-modes accessible to an interferometer are determined by the shortest baselines.
Therefore, a compacted radio interferometer array is required for probing the cosmic LSS, especially for the scales of
baryon acoustic oscillation (BAO). To date, several interferometers, 
such as the Canadian Hydrogen Intensity Mapping Experiment \citep[CHIME,][]{Bandura:2014gwa}, 
Hydrogen Intensity and Real-time Analysis eXperiment \citep[HIRAX,][]{Newburgh:2016mwi} and 
Tianlai \citep{Chen:2012xu,Wu:2020jwm}, are designed as compacted arrays to probe the BAO. 

The MeerKAT radio telescope array with 64 dish antennas of 13.5 m diameter is located in the Northern Cape Province
of South Africa. As a precursor of SKA, MeerKAT has already been operating and producing preliminary results
\citep{Pourtsidou:2017era,Wang:2020lkn,Knowles:2020cuc,TernideGregory:2021tus,deVilliers:2022vhg}. 
The MeerKAT Large Area Synoptic Survey 
\citep[MeerKLASS,][]{MeerKLASS:2017vgf,Wang:2020lkn,2021MNRAS.501.4344L,Irfan:2021xuk} is
proposed for cosmological studies with single-dish-mode \hi IM. 
Using the MeerKLASS pilot survey data, \citet{2022arXiv220601579C} reported the cross-correlation 
power spectrum detection. 
In addition, the small field deep surveys using MeerKAT interferometer mode, 
such as the MeerKAT International GHz Tiered Extragalactic Exploration 
\citep[MIGHTEE,][]{Jarvis:2017aml,Paul:2020ank,refId0}, can also provide a \hi\ cube. 
With such \hi\ cube, \citet{2023arXiv230111943P} reported the \hi IM auto power spectrum detection 
on Mpc scales. 
In this paper, we investigate the performance of MeerKAT interferometer-mode \hi IM survey in measuring 
the power spectrum with different survey strategies and forecast the constraints on 
cosmological parameters in typical dark energy models.

We only consider the most typical dark energy models in this work. Among all the dark energy models, the $\Lambda$ cold dark matter ($\Lambda$CDM) model is the simplest. It is though simple but extremely important in cosmology because it can well explain almost all the current cosmological observations. In the $\Lambda$CDM model, dark energy is assumed to be a cosmological constant $\Lambda$ with the equation of state $w=-1$, and thus the dark-energy density remains a constant during the cosmological evolution. Note here that the equation of state of dark energy is defined as $w=p_{\rm de}/\rho_{\rm de}$, with $p_{\rm de}$ and $\rho_{\rm de}$ the pressure and density of dark energy, respectively. Although the $\Lambda$CDM model is widely viewed as a standard model of cosmology, it still faces huge challenges in both theoretical and observational aspects. In the theoretical aspect, the cosmological constant has always been suffering with the ``fine-tuning'' and ``cosmic coincidence'' problems \citep{Zlatev:1998tr,Sahni:1999gb,2005astro.ph.10059B}. In the observational aspect, the so-called ``Hubble tension'' problem has been widely viewed as triggering a new crisis in cosmology, known as the ``Hubble crisis'' \citep{Verde:2019ivm,Riess:2019qba,DiValentino:2021izs,Poulin:2018cxd,Guo:2018ans,Guo:2019dui,Cai:2021wgv,Vagnozzi:2019ezj,Perivolaropoulos:2021jda,Gao:2021xnk,2022arXiv221213146G,Zhang:2019cww,Liu:2019awo,Ding:2019mmw}. Therefore, a dynamical dark energy was proposed very early on. Actually, in order to explore and understand the nature of dark energy, the first step should be precisely measuring the equation of state of dark energy that governs how dark-energy density evolves with the cosmic expansion. 

The most simplest dynamical dark energy model assumes that $w$ is a constant. For convenience, we call this model the $w$CDM model in this work. This model is very useful in exploring the nature of dark energy with observational data due to its simplicity. However, generally speaking, the equation of state of dark energy should be time-varying, and thus $w$ should be a function of the redshift $z$ (or equivalently, the scale factor of the Universe, $a$).

The two-parameter form of the parametrization of dark-energy equation of state, $w(a)=w_0+(1-a)w_a$ \citep{Chevallier:2000qy,Linder:2002et}, has been widely used in the exploration of the nature of dark energy with cosmological observations, since it can cover many possible $w(z)$ forms in the late Universe. For convenience, this parametrization model is called the $w_0w_a$CDM model in this work.

In this paper, we discuss the cosmological constraints only in the most typical dark energy models, i.e., the $\Lambda$CDM, $w$CDM, and $w_0w_a$CDM models. In these models, the late-time expansion history of the Universe is described by the following equation,
\begin{equation}
    H(z)=H_0\sqrt{\Omega_{\rm m}(1+z)^3+(1-\Omega_{\rm m})f_{\rm de}(z)},
\end{equation}
where $H_0$ is the Hubble constant and $\Omega_{\rm m}$ is the present-day fractional density of matter. Here, the function $f_{\rm de}(z)$ describes the cosmological evolution of the dark-energy density. For $\Lambda$CDM, $f_{\rm de}(z)=1$; for $w$CDM, $f_{\rm de}(z)=(1+z)^{3(1+w)}$; for $w_0w_a$CDM, $f_{\rm de}(z)=(1 + z) ^{3(1+w_0+w_a)} \exp{(-3w_a z/(1+z))}$.

%In this work, it is worth mentioning that we employ a novel approach, i.e. the `delay spectrum' analysis, which is first applied in the PAPER observation \citep{Parsons:2009ju}. 
%In the actual \hi\ IM observation, there are many other astrophysical processes, such as synchrotron and free-free emission from our own Galaxy and extra-galactic point sources, which emit radiation in the same frequency ranges \citep{DiMatteo:2001gg,DiMatteo:2004jha}. 
In this work, we employ the ``delay spectrum'' approach, which is first applied in the observation of
Precision Array for Probing the Epoch of Reionization \citep[PAPER,][]{Parsons:2009ju}. 
At present, it has been successfully applied to \hi IM surveys \citep{2023arXiv230111943P}. 
With such an approach, we can deal with visibilities directly, which are primary data products of radio interferometers. 
By Fourier transformation of the visibilities across the frequencies, we can effectively capture signal delay between antenna pairs.
It is known that the bright foreground emissions contaminate the cosmic \hi signal in the same frequency ranges, such as 
the synchrotron and free-free emission from the Galaxy and extra-galactic point sources.
Generally, the foreground contamination components have smooth frequency spectra, which can be separated 
from the cosmic \hi fluctuation in the ``delay spectrum'' space
\citep{Parsons:2012qh,Liu:2014bba,Liu:2014yxa,Liu:2019awk}.

This paper is organized as follows. In Section~\ref{sec:method}, we provide a detailed description of estimating \hi\ signal power spectrum, MeerKAT survey strategy and its system noise, foregrounds and shot noise. In Section~\ref{sec:results}, we present the constraint results of the power spectrum and cosmological parameters. Finally, the conclusion is given in Section~\ref{sec:conclusion}. In our simulation, we assume a flat $\Lambda$CDM model as the fiducial model and keep the fiducal values of cosmological parameters fixed to \emph{Planck} 2018 results \citep{Planck:2018vyg}.

\section{Methodology}\label{sec:method}

\subsection{\hi\ delay spectrum}\label{sec:delayspectrum}

%\hi\ IM observation directly measures \hi\ brightness temperature. The sky brightness temperature can be defined as 
The \hi brightness temperature fluctuation across the survey volume is expressed as,
\begin{equation}
T_{\rm b}(\bm{\theta}, \nu) = \bar{T}_{\rm b}(\nu) + \delta T_{\rm b}(\bm{\theta}, \nu) \ ,  
\label{Temp}
\end{equation}
where $\bm{\theta}$ is the position vector on the sky, $\nu$ is the observation frequency, 
$\bar{T}_{\rm b}(\nu)$ and $\delta T_{\rm b}(\bm{\theta}, \nu)$ denote the isotropic and 
fluctuating components of the 
\hi\ brightness temperature distribution, respectively.

Radio interferometer observation produces the \hi\ signals visibilities, which are the cross-correlation signals between 
each pair of antennas. Assuming the flat-sky approximation, the visibility for a pair of antennas is given by
\begin{equation}
V(\bm{u}, \nu) = \int A(\bm{\theta}, \nu) \delta T_{\rm b}({\bm{\theta}}, \nu) e^{-i2\pi\bm{u}\cdot\bm{\theta}} {\rm} {\rm d}\Omega \ , 
\label{Vi}
\end{equation}
where $A(\bm{\theta}, \nu)$ denotes the primary beam response of the telescope in the direction of $\bm{\theta}$,
${\rm d}\Omega$ represents the solid angle element. 
$\bm{u} = \nu\bm{b}/c$ is defined as the baseline vector, where $\bm{b}$ is the baseline vector between a pair of antennas 
and $c$ is the speed of light. The visibility function can be Fourier transformed to the `delay spectrum' space via
\begin{equation}
\tilde{V}(\bm{u}, \tau) = \int V(\bm{u}, \nu) e^{-i2\pi\nu\tau}{\rm d}\nu,
\end{equation}
where $\tau=1/\delta \nu$ is the corresponding delay of frequency interval $\delta \nu$.
Following \cite{McQuinn:2005hk}, \cite{Parsons:2012qh} and \cite{Liu:2019awk}, 
the \hi\ power spectrum is expressed in the form of `delay spectrum',
\begin{equation}
P_{\rm{D}}(k_\perp, k_\parallel) \equiv \frac{A_e}{\lambda^2 B} 
\frac{r^2 r_{\nu}}{B} \left|\tilde{V}(\bm{u},\tau)\right|^2 
\left( \frac{\lambda^2}{2k_{\rm{B}}} \right) ^2,
\label{P_D}
\end{equation}
where $A_e$ and $B$ are the effective antenna area and bandwidth, respectively, 
$\lambda$ denotes the wavelength at the centre of the band, 
$r$ is the comoving distance to the redshift $z$ corresponding to $\lambda$,
$r_{\nu}$ is the comoving width along the line-of-sight (LoS) corresponding to the redshift range determined by $B$,
and $k_{\rm{B}}$ is the Boltzmann constant.
Here, $k_\perp$ and $k_\parallel$ are the Fourier wave vectors perpendicular and parallel
to the LoS, respectively. They are related to the 
interferometer variables via
\begin{equation}
k_\perp = \frac{2\pi|\bm{u}|}{r},\quad 
k_\parallel = \frac{2\pi\tau\nu_{21}H(z)}{c(1+z)^2}.\label{eq:kk}
\end{equation}
where $\nu_{21} = 1420~\rm{MHz}$ is the rest-frame frequency of the 21 cm line. $H(z)$ denotes the Hubble parameter as a function of redshift $z$.

There are various advantages of this `delay spectrum' method \citep{Parsons:2012qh,Vedantham:2011mh,Paul:2016blh}. 
The different spectral behaviours between \hi\ signal and foreground make it possible to isolate the latter in the Fourier space. 
In addition, the Fourier conjugate variable is associated with the LoS cosmological distance, therefore the ‘delay spectrum’ 
constructed in this method can recover the cosmological 3D \hi\ power spectrum \citep{Parsons:2012qh,Liu:2014bba,Liu:2014yxa}.

\subsection{\hi\ signal power spectrum}\label{sec:signal}

The mean sky brightness temperature of \hi\ 21 cm emission is expressed as
\citep{Santos:2015gra,MeerKLASS:2017vgf}
\begin{equation}
\bar{T}_{\rm b}(z) \approx 566h \left(\frac{H_0}{H(z)}\right) \left(\frac{\Omega_{\rm{\hi}}(z)}{0.003}\right)(1+z)^2  \ \mu\rm{K} \ ,
\end{equation}
where $H_0 = 100 h$ km s$^{-1}$ and 
%$H(z)$ denotes the Hubble parameter as a function of redshift $z$. 
$\Omega_{\rm{\hi}}(z)$ is the fractional density of \hi, which can be written as 
\begin{equation}
\Omega_{\rm{\hi}}(z) = \frac{\rho_{\rm{\hi}}(z)}{\rho_{c,0}}(1+z)^{-3} \ ,
\end{equation}
where $\rho_{c,0}$ is the critical density today and the proper \hi\ density is calculated by
\begin{equation}
\rho_{\rm{\hi}}(z) = \int_{M_{\rm min}}^{M_{\rm max}}{\rm d}M \frac{{\rm d}n}{{\rm d}M}(M,z) M_{\rm{\hi}}(M,z) \ ,
\end{equation}
where $M$ denotes the dark matter halo mass, ${{\rm d}n}/{{\rm d}M}$ is the proper halo mass 
function and $M_{\rm{\hi}}(M,z)$ denotes the \hi\ mass in a halo of mass $M$ at redshift $z$.
Throughout this paper, we assume a simple power-law model following \cite{Santos:2015gra}, 
i.e., $M_{\rm{\hi}}(M) = A M^\alpha$ with $A\approx 220$ and $\alpha=0.6$ that can fit both 
low- and high-redshift observations within reasonable accuracy.

Considering the redshift space distortion (RSD) effect \citep{Kaiser:1987qv}, 
the \hi\ power spectrum is expressed as
\begin{equation}
P_{\rm{\hi}}(k, \mu, z) =  \bar{T}^2_{\rm b}(z) F_{\rm{RSD}}(k, \mu) P(k, z) \ , \label{Pk}
\end{equation}
where $\mu \equiv k_{\parallel}/k$. The matter power spectrum $P(k, z) = D^2(z) P(k, z=0)$,
where $D(z)$ is the growth factor and $P(k, z=0)$ is the matter power spectrum at z = 0.
$P(k, z=0)$ is obtained with \texttt{CAMB} \citep{Lewis:1999bs}. 
$F_{\rm{RSD}}(k, \mu)$ represents the RSD effect which is expressed as
\begin{equation}
F_{\rm{RSD}}(k, \mu) = \left( b^2_{\rm{\hi}}(z) + f \mu^2 \right) ^2 {\rm{exp}} \left( -k^2 \mu^2 \sigma^2_{\rm{NL}} \right), \label{RSD}
\end{equation}
where $b_{\rm{\hi}}(z)$ is the \hi\ bias, written as
\begin{equation}
b_{\rm{\hi}}(z) = \rho_{\rm{\hi}}^{-1}(z) \int_{M_{\rm min}}^{M_{\rm max}} {\rm d}M \frac{{\rm d}n}{{\rm d}M} M_{\rm{\hi}}(M, z) b(M, z) , \label{bias_HI}
\end{equation}
where $b(M, z)$ is the halo bias,
$f\equiv {\rm d} {\rm ln}D / {\rm d} {\rm ln}a$ is the linear growth rate 
with $a$ being the scale factor, $\sigma_{\rm{NL}}$ is the nonlinear dispersion scale 
with the value of $\sigma_{\rm{NL}}=7~\rm{Mpc}$. 
In this paper, we simply employ the fitting functions of $\Omega_{\rm{\hi}}(z)$ and $b_{\rm{\hi}}(z)$ 
%for convenience, $\Omega_{\rm{\hi}}(z)$ and $b_{\rm{\hi}}(z)$ are employed with the fitting functions 
following \citet{MeerKLASS:2017vgf}, 
\begin{align}
\Omega_{\rm{\hi}}(z) &=  4.8 \times 10^{-4} + 3.9 \times 10^{-4} z + 6.5 \times 10^{-5} z^2 , \\
b_{\rm{\hi}}(z) &=  0.67 + 0.18 z + 0.05 z^2. 
\end{align}

\subsection{MeerKAT noise power spectrum}\label{sec:noise}

\begin{figure}
	\centering
  	\begin{minipage}{0.95\linewidth}
	\centering
  	    \includegraphics[width=\textwidth]{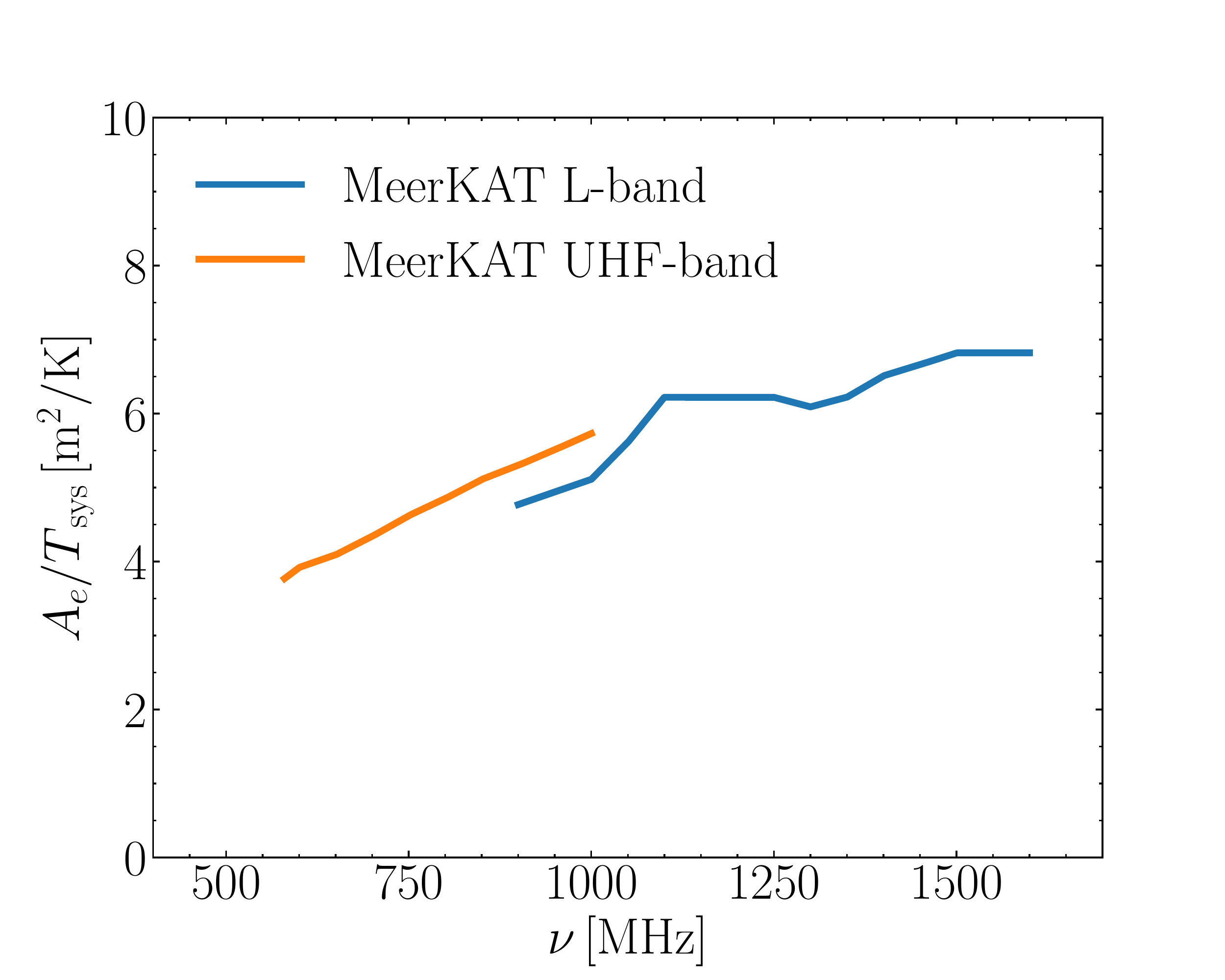}
  	\end{minipage}
    \caption{
    The sensitivity designs for MeerKAT receivers, shown as $A_e/T_{\rm{sys}}$ for L-band and UHF-band. 
    }
    \label{AT}
\end{figure}

\begin{figure*}
	\centering
	\begin{minipage}{0.45\linewidth}
	\centering
  	    \includegraphics[width=\textwidth]{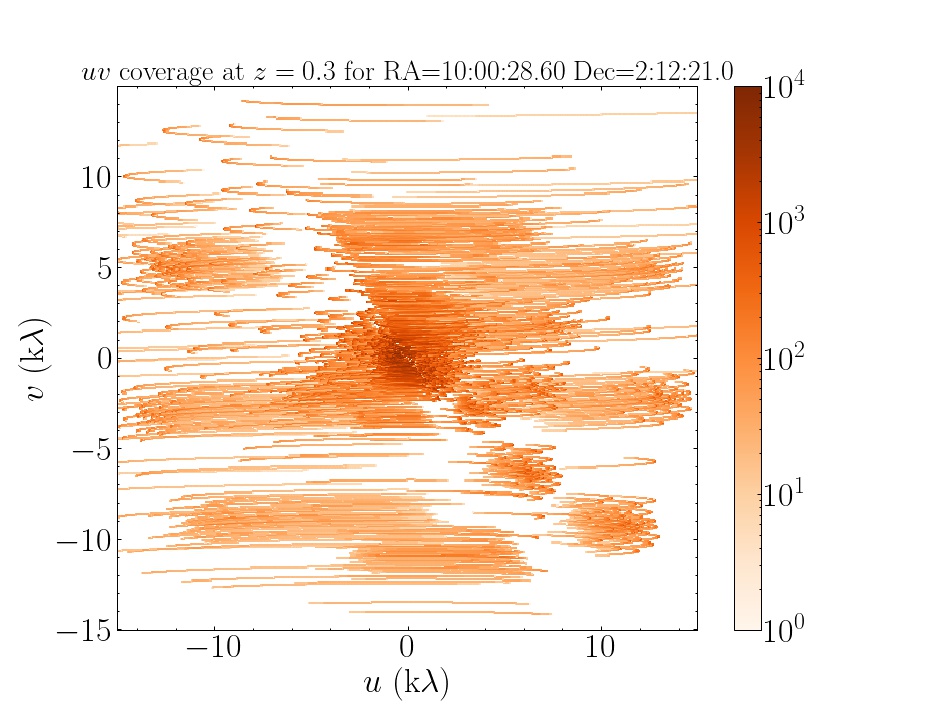}
  	\end{minipage}
  	\begin{minipage}{0.45\linewidth}
	\centering
  	    \includegraphics[width=\textwidth]{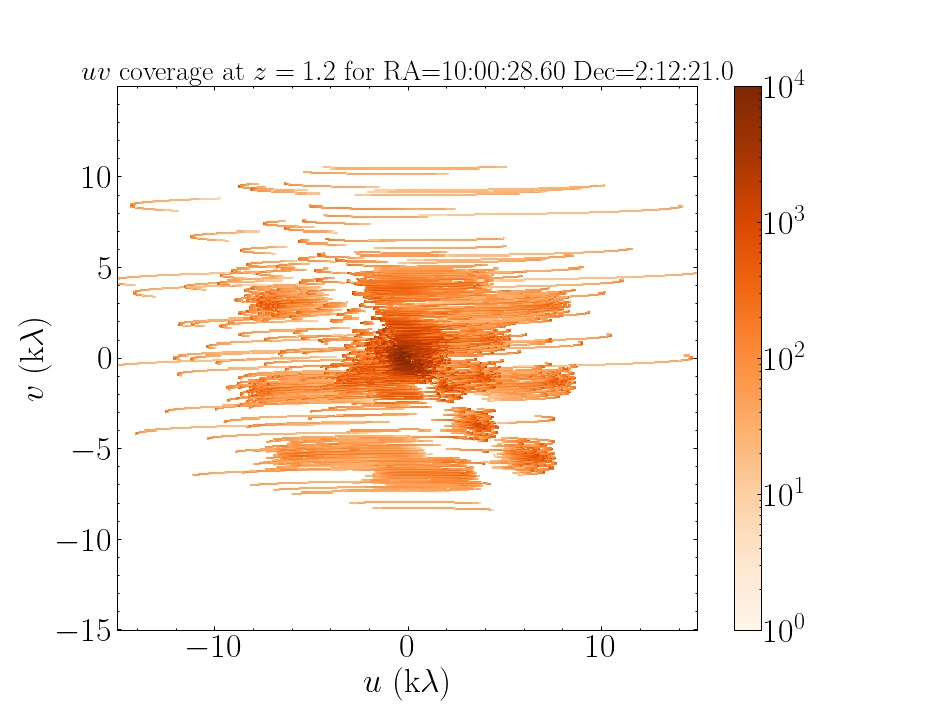}
  	\end{minipage}
    \caption{The distribution of MeerKAT baselines on a two-dimensional (2D) $uv$ plane for $10$ h 
    tracking of the COSMOS field with sub-bands in L-band 
    centering at $z=0.3$ (left panel) and in UHF-band centering at $z=1.2$
    (right panel). 
    The $uv$ plane is segmented onto a discrete grid with cell-size 
    $\Delta u = \Delta v = 60 \lambda$. 
    The color signifies the number of $uv$ points on the grid. 
    }
    \label{fig:uv_2d}
\end{figure*}

\begin{figure*}
	\centering
	\begin{minipage}{0.43\linewidth}
	\centering
  	    \includegraphics[width=\textwidth]{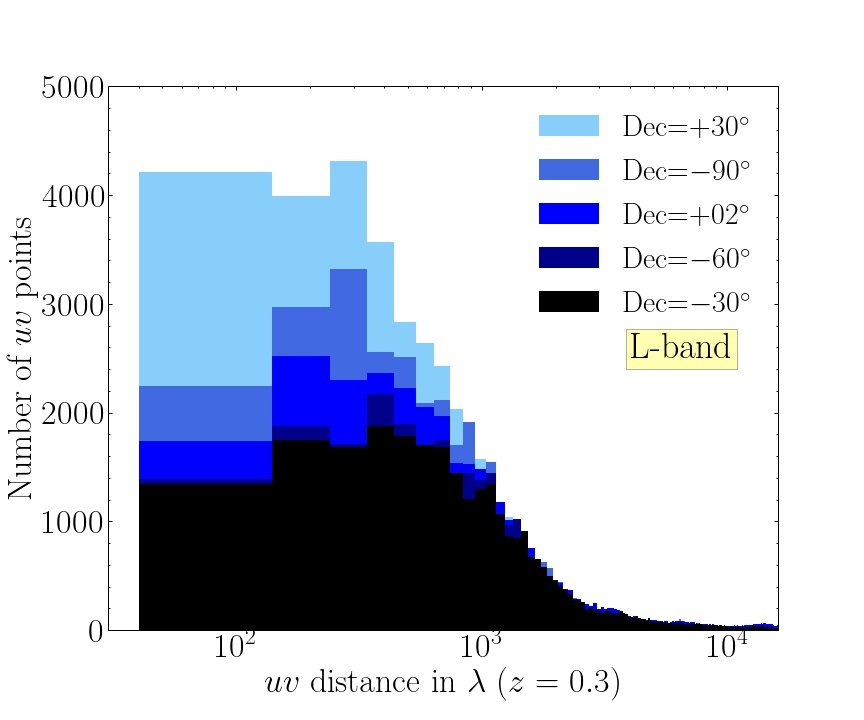}
    \end{minipage}
    \begin{minipage}{0.43\linewidth}
    \centering
        \includegraphics[width=\textwidth]{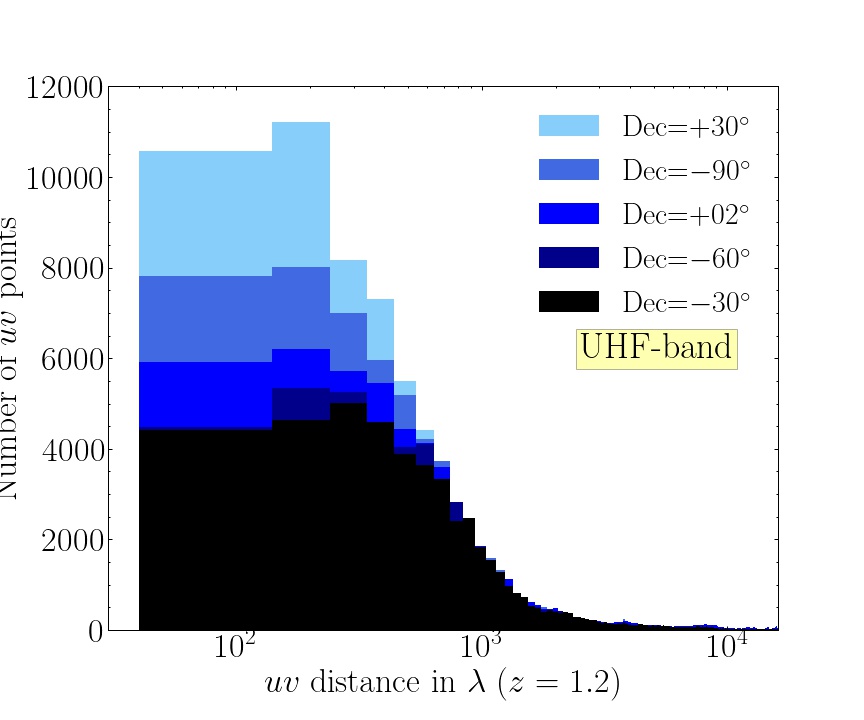}
    \end{minipage}
    \caption{The average number of MeerKAT baselines as a function of $uv$ distance, 
    $|\bm{u}| = \sqrt{u^2 + v^2}$, with bin size of $\Delta |\bm{u}| = 100 \lambda$,
    for $10$ h tracking with sub-bands in L-band centering at $z=0.3$ (left panel) and in UHF-band centering at $z=1.2$ (right panel). }
    \label{fig:uv_1d}
\end{figure*}

The total thermal noise power spectrum can be written as \citep{Bull:2014rha}
\begin{equation}
P_{\rm{N}}(k, \mu, z) = r^2(z) r_{\nu}(z) \frac{T^2_{\rm{sys}} \lambda^4}{n_{\rm{pol}} \nu_{21} t_{\rm{int}} A_e^2 n(\bm{u})} \ , \label{Pnoise}
\end{equation}
where %$r_{\nu}(z)$ is the radial comoving distance. 
$n_{\rm{pol}} = 2$ represents the dual-polarization of the MeerKAT instrument %$\nu_{21} = 1420\, \rm{MHz}$ is the rest-frame frequency of the 21 cm line. 
and $t_{\rm{int}}$ is the integration time. The ratio of MeerKAT effective antenna area and system temperature,
$A_e/T_{\rm{sys}}$, is frequency dependent. Currently, there are two frequency bands available for 
observation, i.e., the L-band ($900$--$1700$ MHz) and the UHF-band ($580$--$1000$ MHz). 
Because of the serious RFI contamination in the L-band frequency range, only the frequency range of 
$900$--$1200$ MHz ($0.18<z<0.58$) is used in our analysis. 
The full UHF-band is used in this analysis, corresponding to $0.42<z<1.45$.  
In Fig.~\ref{AT}, we show $A_e/T_{\rm{sys}}$ for L-band and UHF-band. 
These values are obtained from the MeerKAT website.\footnote{\href{http://public.ska.ac.za/meerkat/meerkat-schedule}{http://public.ska.ac.za/meerkat/meerkat-schedule}}

In Eq.~(\ref{Pnoise}), $n(\bm{u})$ is the baseline density referring to the detailed $uv$ coverage of 
a particular observation. We employ the actual MeerKAT antenna coordinates and assume tracking the 
COSMOS field (RA=10h01m, Dec=+02d12m) following \citet{Paul:2020ank}. 
Particularly, $n(\bm{u})$ is also a function of frequency. We divide the full frequency range into 
dozen $\Delta\nu=60$ MHz sub-bands. The $uv$ coverage is assumed to be uniform within each sub-band 
and simulated according to the center frequency of each sub-band.
In Fig.~\ref{fig:uv_2d}, we show the simulated $uv$ coverage corresponding to one of the sub-bands
in the L-band centering at $z=0.3$ and one in UHF-band centering at $z=1.2$ in the left and right panels,
respectively. For both cases, we assume $10$ h tracking observation of the COSMOS field spanning over 
two days (the start time is 14:15 and 13:33 at LST, respectively). 
The $uv$ plane is segmented onto a discrete grid with cell size $\Delta u= \Delta v = 60 \lambda$. 
The color represents the number of $uv$ points within the grid. 
It is clear that, in the short $uv$ distance region, there are more $uv$ samples at the 
low-frequency band than at the high-frequency band.

Because of the uniform $uv$ coverage assumption across the sub-bands, 
%the baseline density $n(\bm{u})$ and the corresponding total thermal noise power spectrum $P_{\rm N}$ are only the functions of $k_\perp$. 
in the same sub-band, the total thermal noise power spectrum $P_{\rm N}$ is mainly determined by the baseline density $n(\bm{u})$ that is related to $k_\perp$.
According to Eq.~(\ref{eq:kk}), $k_\perp$ is proportional to the $uv$ distance, i.e. $|\bm{u}| = \sqrt{u^2 + v^2}$. The circular averaged $uv$ coverage within a $|\bm{u}|$ shell of width $\Delta |\bm{u}| = 100 \lambda$ are shown in Fig.~\ref{fig:uv_1d}, where the left panel shows the distribution corresponding to the sub-band centering at $z=0.3$ and the right panel shows the one centering at $z=1.2$. Since $|k_\perp|$ is proportional to the $uv$ distance, the more densely populated $uv$ points at smaller distances mean the higher sensitivity at the smaller $|k_\perp|$ modes. 

It is known that the $uv$ coverage also depends on the pointing direction. In order to investigate the influence of the different sky zones, we also show in Fig.~\ref{fig:uv_1d} the numbers of $uv$ points with 10 h tracking at different declinations: $\rm{Dec}=+30^\circ, +02^\circ, -30^\circ, -60^\circ, -90^\circ$. The case of $\rm{Dec}=+02^\circ$ is the same as tracking the COSMOS field and the case of $\rm{Dec}=-30^\circ$ corresponds to tracking a field that the transit line passes near zenith of the MeerKAT site. 
It is obvious that when the field is targeted farther from the zenith, the number of short baselines rises substantially for both the L-band and the UHF-band, which potentially increases the sensitivity at the smaller $|k_\perp|$ modes.

\subsection{The foreground wedge and shot noise}\label{sec:foreground}

The foreground contamination, which is several orders of magnitude stronger than \hi\ signal, is the major challenge in recovering the \hi\ LSS. 
Since the foreground spectrum is smooth across frequency channels, 
it only contaminates the power spectrum close to the smallest $k_\parallel$. 
Therefore, we exclude the $k_\perp$--$k_\parallel$ space modes within the 
foreground wedge \citep{Datta:2010pk,Morales:2012kf,Liu:2014bba,Liu:2014yxa,Pober:2014lva,Seo:2015aza} 
that can be expressed as
\begin{equation}
k_\parallel < \frac{r(z) H(z) \sin({\theta})}{c(1+z)} k_\perp \ ,  \label{wedge_eq}
\end{equation}
where $\theta$ denotes the field of view of the interferometer. The foregrounds only play a dominant role within the lower-end of $k_\parallel$ and the foreground wedge region. 
In our analysis, the foreground contaminated region is masked. 
%Thus, we mask the foreground wedge region in our analysis. 
However, in real observation, instrumental issues, such as beam response and polarization leakage effect, can exacerbate foreground contamination. 
It is inevitable to cause foreground leakage into the high-$k_\parallel$ modes, 
even though deep learning methods can help solve this problem to some extent \citep{2022ApJ...934...83N,2022arXiv221208773G}. 
More recently, \citet{2023arXiv230111943P} presented a groundbreaking approach to \hi IM by employing a foreground avoidance method. They successfully suppressed foreground power leakage to higher $k_\parallel$ modes by multiplying the visibility function with the Blackman-Harris spectral window function. 
In this work, we simplify this process and assume that foreground leakage can be completely avoided.

In addition, shot noise needs to be taken into account in \hi\ IM survey. Because of Poisson fluctuations in halo number, the shot noise power spectrum is written as \citep{Bull:2014rha}
\begin{equation}
P_{\rm \hi}^{\rm shot}(z) = \left(\frac{\bar{T}_b (z)}{\rho_{\rm \hi}(z)}\right)^2 
\int_{M_{\rm min}}^{M_{\rm max}}{\rm d}M \frac{{\rm d}n}{{\rm d}M} M_{\rm \hi}^2(M) \ . \label{shot_noise}
\end{equation}
Here \hi\ mass model is consistent with the description in the \hi\ signal power spectrum. 
Since shot noise is very low according to our calculation, it makes a very small contribution to the total noise.

\section{Results}\label{sec:results}

\begin{figure*}
	\centering
	\begin{minipage}{0.45\linewidth}
	\centering
  	    \includegraphics[width=\textwidth]{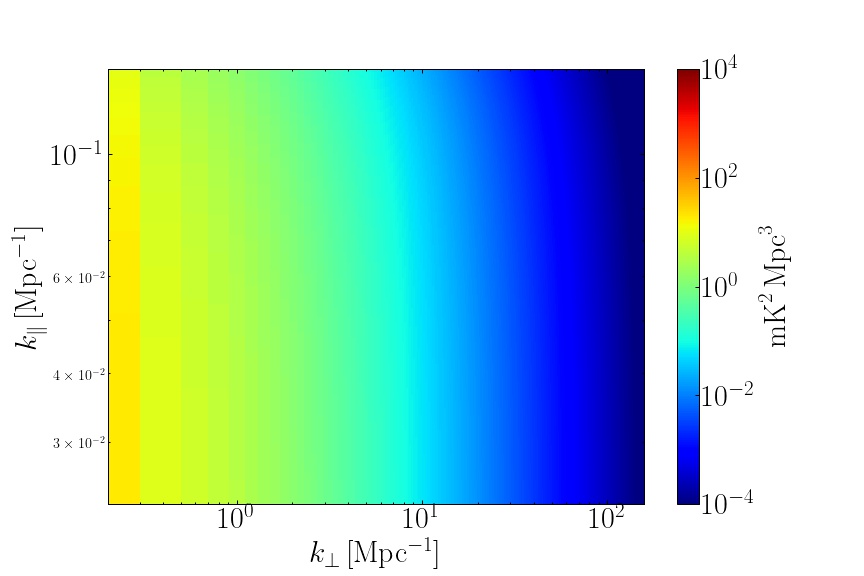}
    \centerline{(a) \hi\ power spectrum}
    \end{minipage}
    \begin{minipage}{0.45\linewidth}
    \centering
        \includegraphics[width=\textwidth]{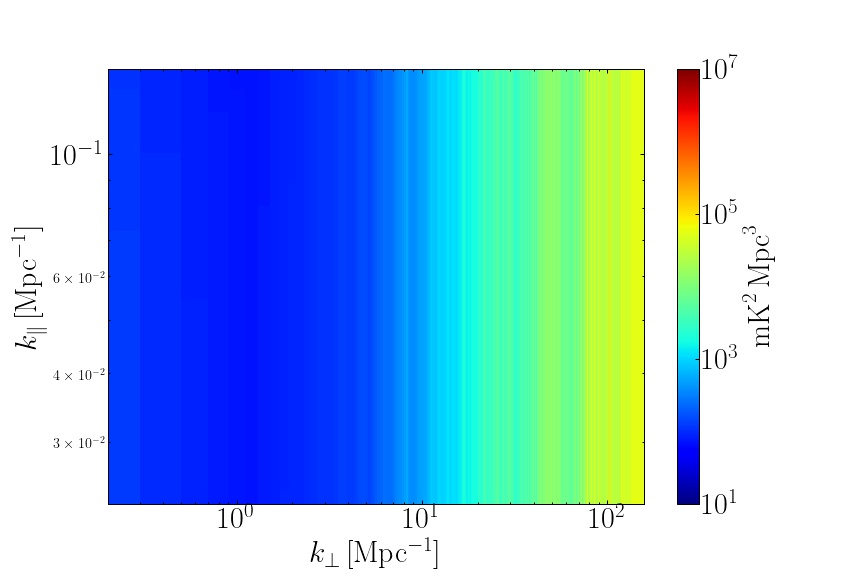}
    \centerline{(b) Total power spectrum}
    \end{minipage}
    %\caption{\textit{Left-hand panel}: 2d \hi\ signal power spectrum $P_{\rm{\hi}}$ at $z=0.3$. \textit{Right-hand panel}: 2d total power spectrum $P_{\rm{tot}}$ with MeerKAT L-band ($z=0.3$) 10 h observations.}
    \caption{2D power spectrum at $z=0.3$. \textit{Left panel}: \hi\ signal power spectrum $P_{\rm{\hi}}$. \textit{Right panel}: Total power spectrum $P_{\rm{tot}}$ with MeerKAT L-band 10 h observation.}
    \label{fig:pk_L}
\end{figure*}

\begin{figure*}
	\centering
  	\begin{minipage}{0.45\linewidth}
	\centering
  	    \includegraphics[width=\textwidth]{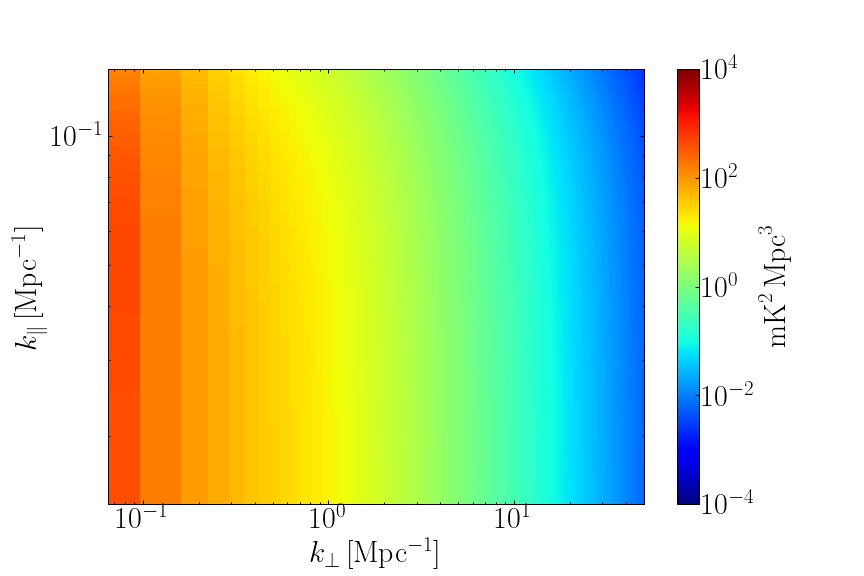}
  	\centerline{(a) \hi\ power spectrum}
  	\end{minipage}
  	\begin{minipage}{0.45\linewidth}
	\centering
  	    \includegraphics[width=\textwidth]{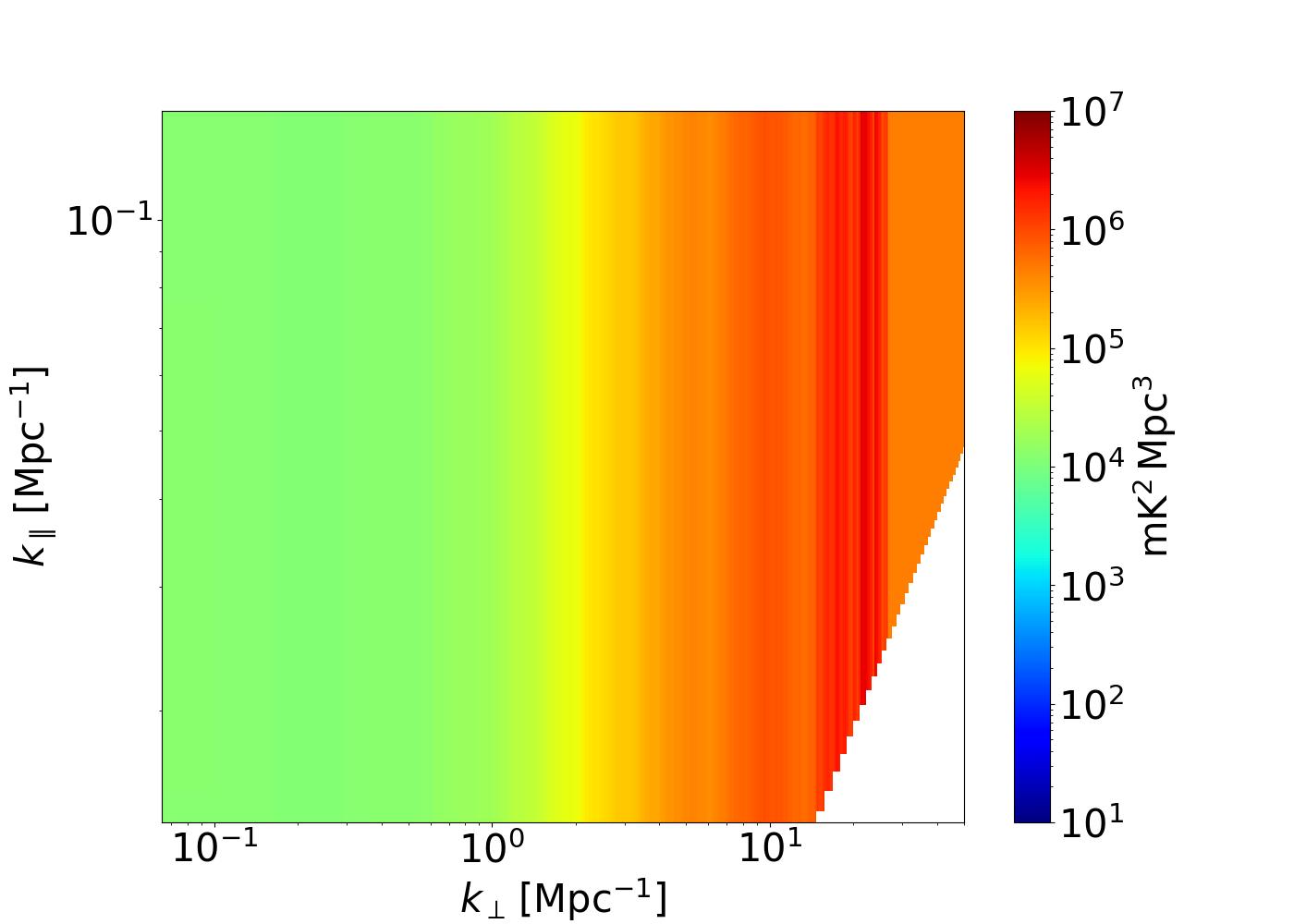}
  	\centerline{(b) Total power spectrum}
  	\end{minipage}
    %\caption{\textit{Left-hand panel}: 2d \hi\ signal power spectrum $P_{\rm{\hi}}$ at $z=1.2$. \textit{Right-hand panel}: 2d total power spectrum $P_{\rm{tot}}$ with MeerKAT UHF-band ($z=1.2$) 10 h observations.}
    \caption{2D power spectrum at $z=1.2$. \textit{Left panel}: \hi\ signal power spectrum $P_{\rm{\hi}}$. \textit{Right panel}: Total power spectrum $P_{\rm{tot}}$ with MeerKAT UHF-band 10 h observation.}
    \label{fig:pk_UHF}
\end{figure*}

In this section, we present the results of \hi\ IM survey analysis. In Section~\ref{powerspectrum}, we give a detailed analysis of the power spectrum in the different survey strategies. The relative errors on $D_A(z)$, $H(z)$ and $f\sigma_8(z)$ and the constraints on cosmological parameters in different dark energy models are showed in Section~\ref{cosmo}.

\subsection{Power spectrum estimation}\label{powerspectrum}

The \hi\ detection is quantified with the relative error of the power spectrum,
\begin{equation}
\left( \frac{\Delta P}{P} \right)^2 = \left[ \frac{1}{8 \pi^2} V_{\rm{bin}} \int k^2 {\rm d}k {\rm d}\mu \left( \frac{P_{\rm{\hi}}(k, \mu)}{P_{\rm{tot}}(k, \mu)} \right)^2 \right]^{-1},  \label{dp/p}
\end{equation}
where $V_{\rm{bin}} = S_{\rm{area}} r^2 r_\nu \frac{\Delta \nu}{\nu_{21}}$ is the survey volume of each redshift bin with the survey area $S_{\rm{area}} = \pi \left( \frac{1}{2}\frac{\lambda}{13.5~{\rm m}} \right)^2 \left(\frac{180}{\pi} \right)^2$. 
$P_{\rm tot}(k, \mu)$ represents the total power spectrum, which consists of the
contributions of \hi signal, thermal noise, and shot noise. 
The effects of foreground contamination are taken into account by 
restricting the integration area in the $k_{\perp}$-$k_{\parallel}$ space.

The signal-to-noise ratio is related to the integration area in the 
$k_{\perp}$-$k_{\parallel}$ space. 
The scales available for \hi\ IM in interferometer mode observation are 
limited by the detailed configuration. In particular, the scale limits for \hi 
IM survey are: 
\begin{align}
k_\parallel^{\rm min} &= 2 \pi/ (r_\nu \Delta \nu / \nu_{21}), \nmsk
k_\parallel^{\rm max} &= 1/\sigma_{\rm NL}, \nmsk
k_\perp^{\rm min} &= 2 \pi |\bm{u}|_{\rm{min}}/r, \nmsk
k_\perp^{\rm max} &= 2 \pi |\bm{u}|_{\rm{max}}/r.
\end{align}
where $|\bm{u}|_{\rm{min}}$ and $|\bm{u}|_{\rm{max}}$ are minimum and maximum interferometer baselines, respectively \citep{Bull:2014rha}. 
As we discussed in Section.~\ref{sec:foreground}, the foreground contamination 
should exist at the lower end of $k_{\parallel}$ and restrict
$k_{\parallel}^{\rm min}$ for \hi IM survey. In our analysis, we assume 
the foreground spectrum to be smooth and only exist in the mode corresponding 
to the largest scale. Thus, the lower bound of $k_{\parallel}$ set by the
foreground contamination overlaps with that set by the survey bandwidth.

In addition, the integration area in the $k_{\perp}$-$k_{\parallel}$ space
also restricted by the foreground wedge, which is defined in Eq.~(\ref{wedge_eq}). 
We show the 2D \hi power spectrum $P_{\rm{\hi}}(k, \mu)$ at $z=0.3$ (in L-band) and
$z=1.2$ (in UHF-band) in the left panels of Figs.~\ref{fig:pk_L} and \ref{fig:pk_UHF},
respectively. The corresponding total power spectra
$P_{\rm tot}(k, \mu)$ are shown in the right panels. 
A clear foreground wedge (the blank area) is shown in the total power spectrum
at $z=1.2$. At lower redshift, the foreground wedge is pushed to larger $k_\perp$, 
which is outside the upper bound set by the MeerKAT baseline configuration.

The power spectrum of the thermal noise is calculated by assuming $10$ h 
tracking observation. The \hi power spectra are totally below the thermal 
noise level for both the L-band and UHF-band. We also assume the 
thermal noise power spectrum can be efficiently removed and achieve unbiased 
\hi power spectrum estimation before proceeding to the cosmology studies. 
In reality, the thermal noise power spectrum can be eliminated by cross-correlating 
the observations of the same survey cube at different sessions
\citep{Switzer:2013ewa,2023arXiv230111943P}.

\begin{figure*}
	\centering
  	\begin{minipage}{0.6\linewidth}
	\centering
  	    \includegraphics[width=\textwidth]{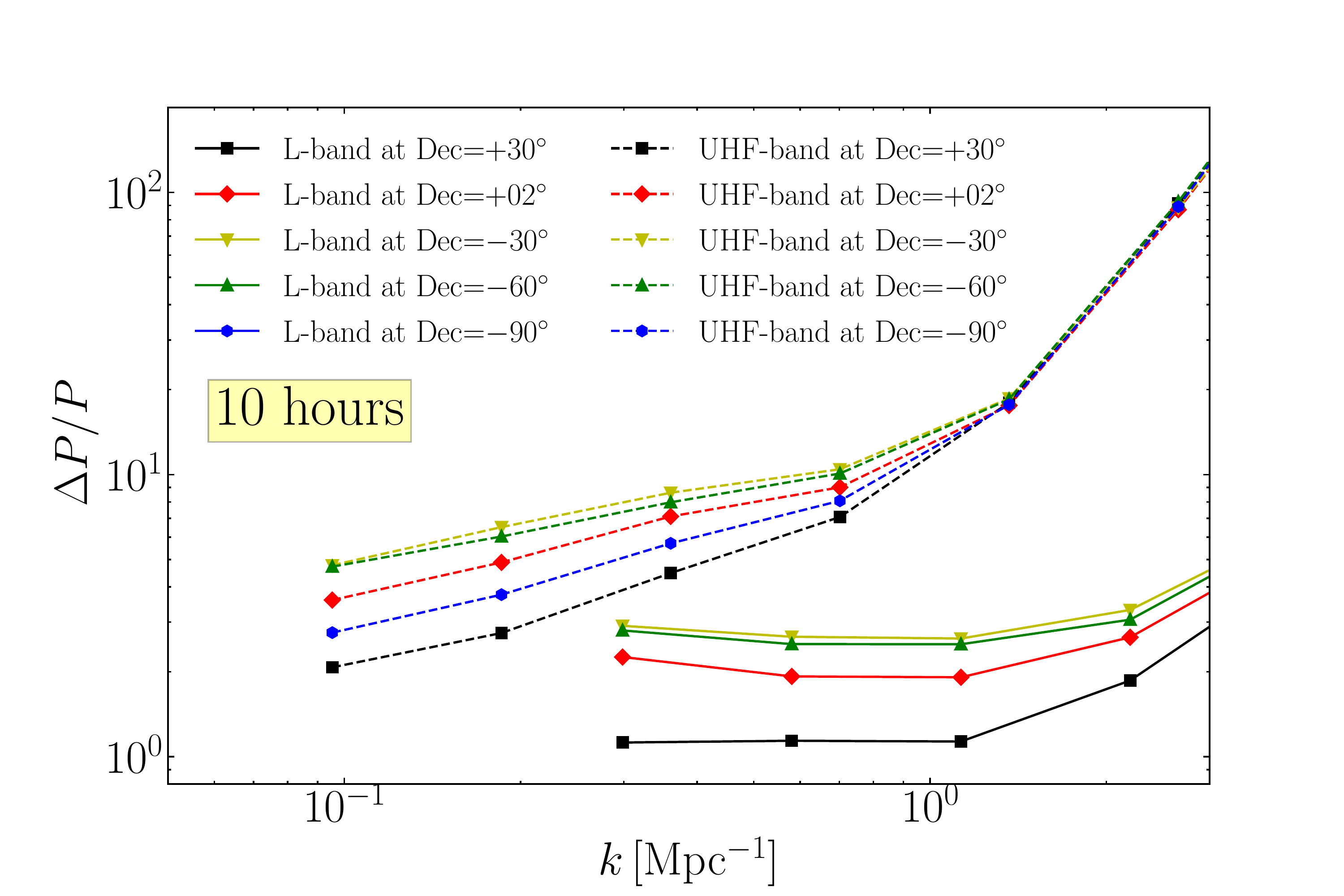}
  	\end{minipage}
  	\begin{minipage}{0.6\linewidth}
	\centering
  	    \includegraphics[width=\textwidth]{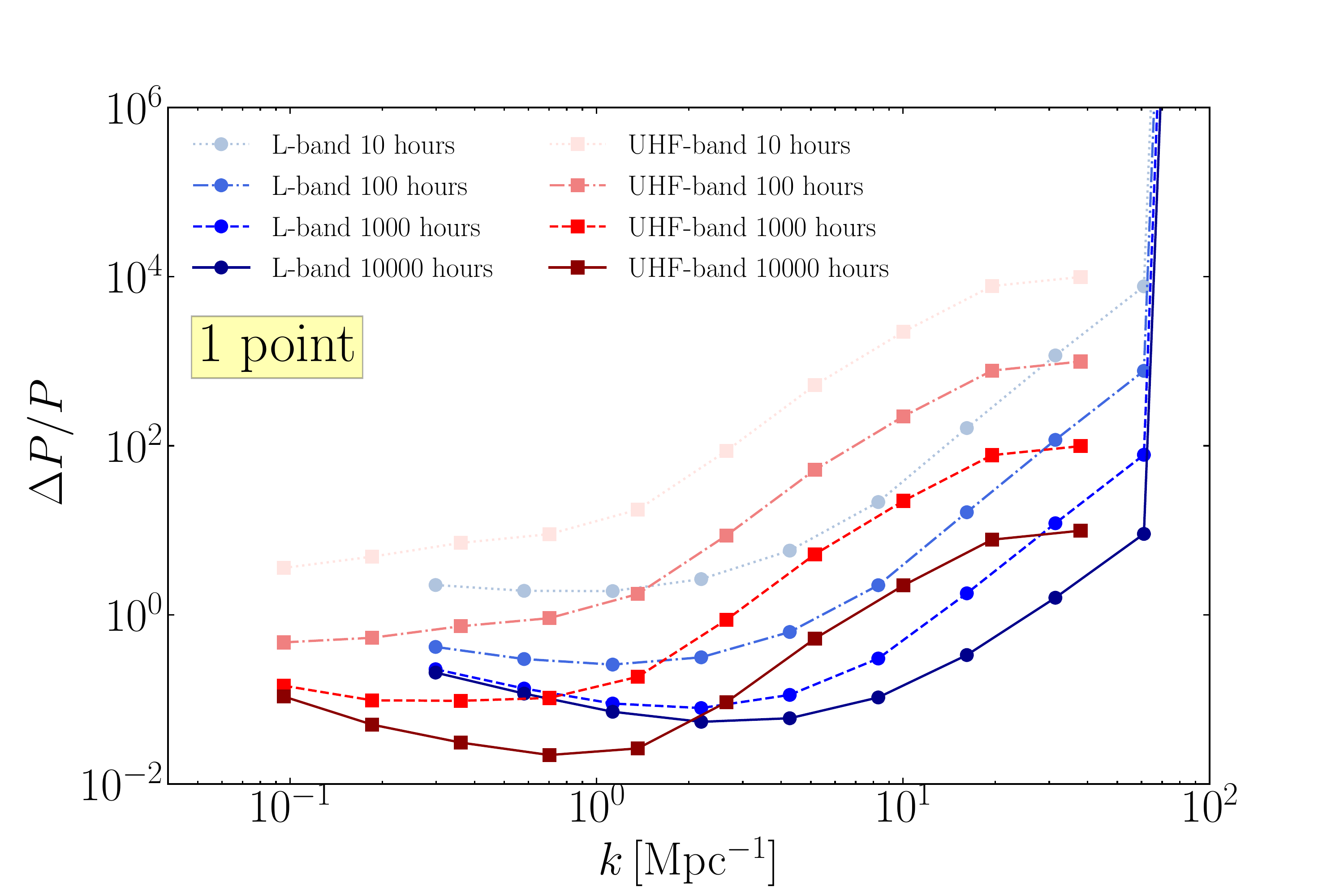}
  	\end{minipage}
        \begin{minipage}{0.6\linewidth}
	\centering
  	    \includegraphics[width=\textwidth]{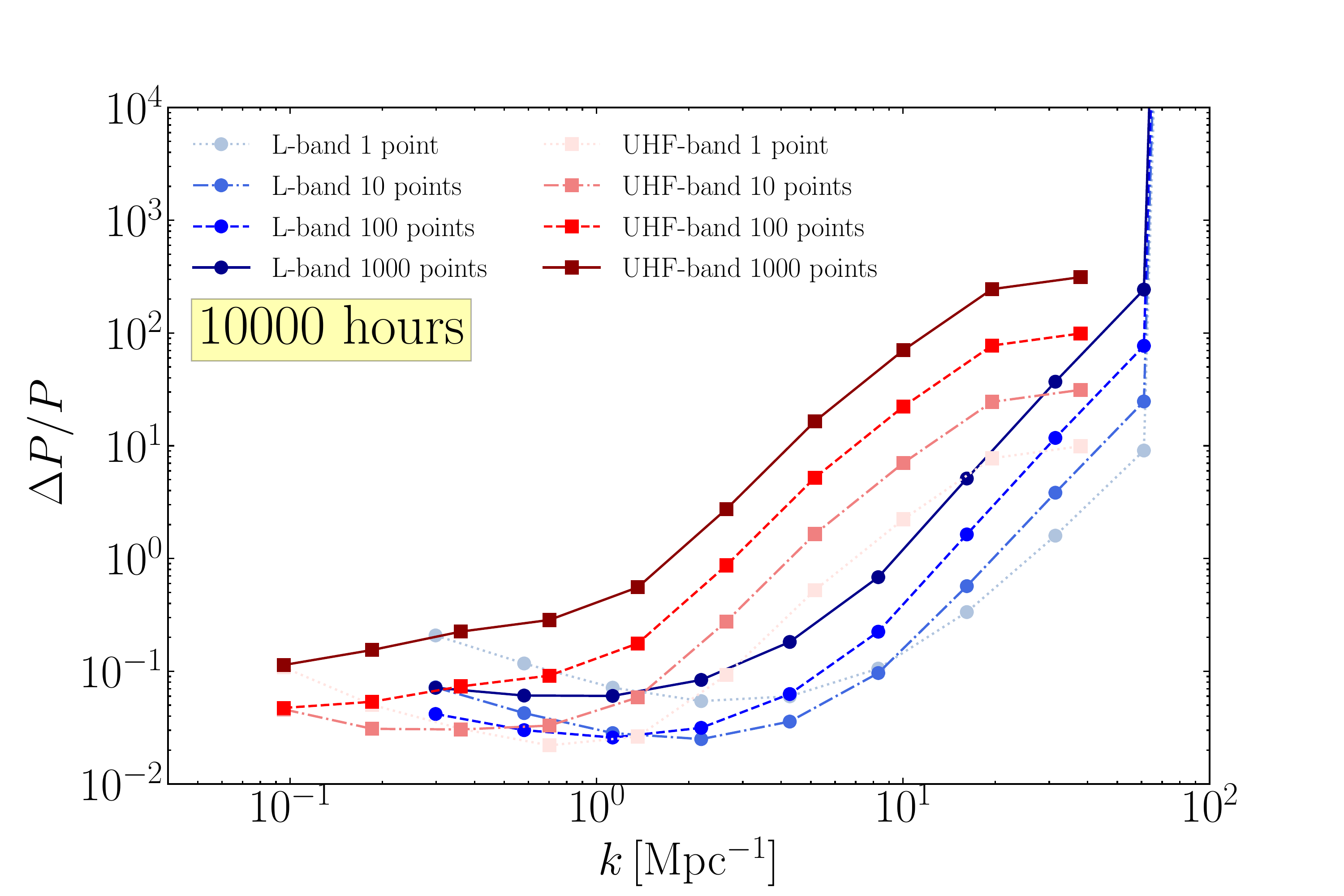}
  	\end{minipage}
    %\caption{\textit{Left-hand panel}: Fractional constraints on $P(k)$ for the different observation time with MeerKAT L-band and UHF-band. \textit{Right-hand panel}: Fractional constraints on $P(k)$ for 10 h observation at the different declination with MeerKAT L-band and UHF-band.}
    %\caption{Fractional constraints on $P(k)$ with MeerKAT L-band and UHF-band. \textit{Left-hand panel}: The different observation time of tracking the COSMOS field. \textit{Right-hand panel}: 10 h observation at the different declination.}
    \caption{Fractional errors on $P(k)$ obtained with MeerKAT L-band and UHF-band. \textit{Top panel}: 10 h observations at the different declinations. \textit{Middle panel}: Different observation times of tracking the COSMOS field. \textit{Bottom panel}: Tracking different numbers of points in a 10000 h observation.}
    \label{fig:pk_error}
\end{figure*}

Firstly, we investigate the influence on the $P(k)$ error when tracking the source at the different declinations. As is shown in Fig.~\ref{fig:uv_1d}, when tracking the source at $\rm{Dec}=+30^\circ, +02^\circ, -30^\circ, -60^\circ, -90^\circ$, completely different numbers of $uv$ points are obtained, e.g. there are more $uv$ points in the shorter $uv$ distance for the case of $\rm{Dec}=+30^\circ$. In the top panel of Fig.~\ref{fig:pk_error}, the relative errors of the power spectrum with different tracking declinations are shown in different colors. For all the cases, we assume a 10 h observation time. The results with L-band and UHF-band are shown in solid and dashed lines, respectively. Here, we divide the whole range of $k$ into $10$ logarithmic bins. It is clear that the power spectrum uncertainty is reduced by more than a factor of two when tracking $\rm{Dec}=+30^\circ$ compared to tracking $\rm{Dec}=-30^\circ$ or $\rm{Dec}=-60^\circ$. The results show that, with limited observation time, the tracking declination has an obvious influence on the results of the constraints on the power spectrum. %With the observational time increasing, when cosmic variance becomes dominating, the impact of the tracking declination will be negligible.

Next, in order to assess the influence of integration time, we increase the integration time by assuming observations on the same field at the same local sidereal time as the existing data on different days, which means that we obtain the same $uv$ points from multiple days coherently to increase the sensitivity of the same $k$ modes. In addition to the current $10$ h observation, we further consider $100$, $1000$ and $10000$ h observations. In the middle panel of Fig.~\ref{fig:pk_error}, we show the fractional errors on the power spectrum $P(k)$ for 10, 100, 1000 and 10000 hours observation of tracking the COSMOS field with MeerKAT L-band (in blue) and UHF-band (in red). The results with different integration times are shown with different color saturations. It can be seen that, compared to L-band, using the UHF-band could measure smaller $k$ modes down to $\sim0.1$ Mpc$^{-1}$, which makes it possible to detect cosmological LSS on larger scales. In addition, it is expected that the lower $\Delta P/P$ can be obtained with the observation time increasing as is shown in the middle panel of Fig.~\ref{fig:pk_error}. With MeerKAT UHF-band 10 h observation, the value of $\Delta P/P$ could reach 1 roughly. The values of $\Delta P/P$ are distinctly reduced when tracking 1000 h, approximately reaching 0.1. However, we find that when the integration time increases from 1000 h to 10000 h, the reduction of $\Delta P/P$ is not significant at low $k$. It is mainly because the cosmic variance, which is limited by the survey volume, plays the dominating role.

Therefore, we consider tracking multiple points equally in the total 10000 h observation. In this case, compared to tracking one point with 10000 h, the survey volume $V_{\rm bin}$ and the thermal noise power spectrum $P_{\rm N}$ are increased by a factor of the number of points $N$. In our analysis, we calculate the additional fractional error on $P(k)$ for $N=10, 100$ and $1000$ in the 10000 h observation, as shown in the bottom panel of Fig.~\ref{fig:pk_error}. In order to constrain cosmological parameters, we expect to obtain lower $\Delta P/P$ in low $k$. We find that the lower values of $\Delta P/P$ in low $k$ are obtained when tracking 100 points for MeerKAT L-band in the total 10000 h observation while tracking 10 points for MeerKAT UHF-band. Therefore, we employ these two survey strategies for MeerKAT L-band and UHF-band, respectively, in the next subsection.

\subsection{Cosmological parameters}\label{cosmo}

\begin{figure*}
	\centering
  	\begin{minipage}{0.3\linewidth}
	\centering
  	    \includegraphics[width=\textwidth]{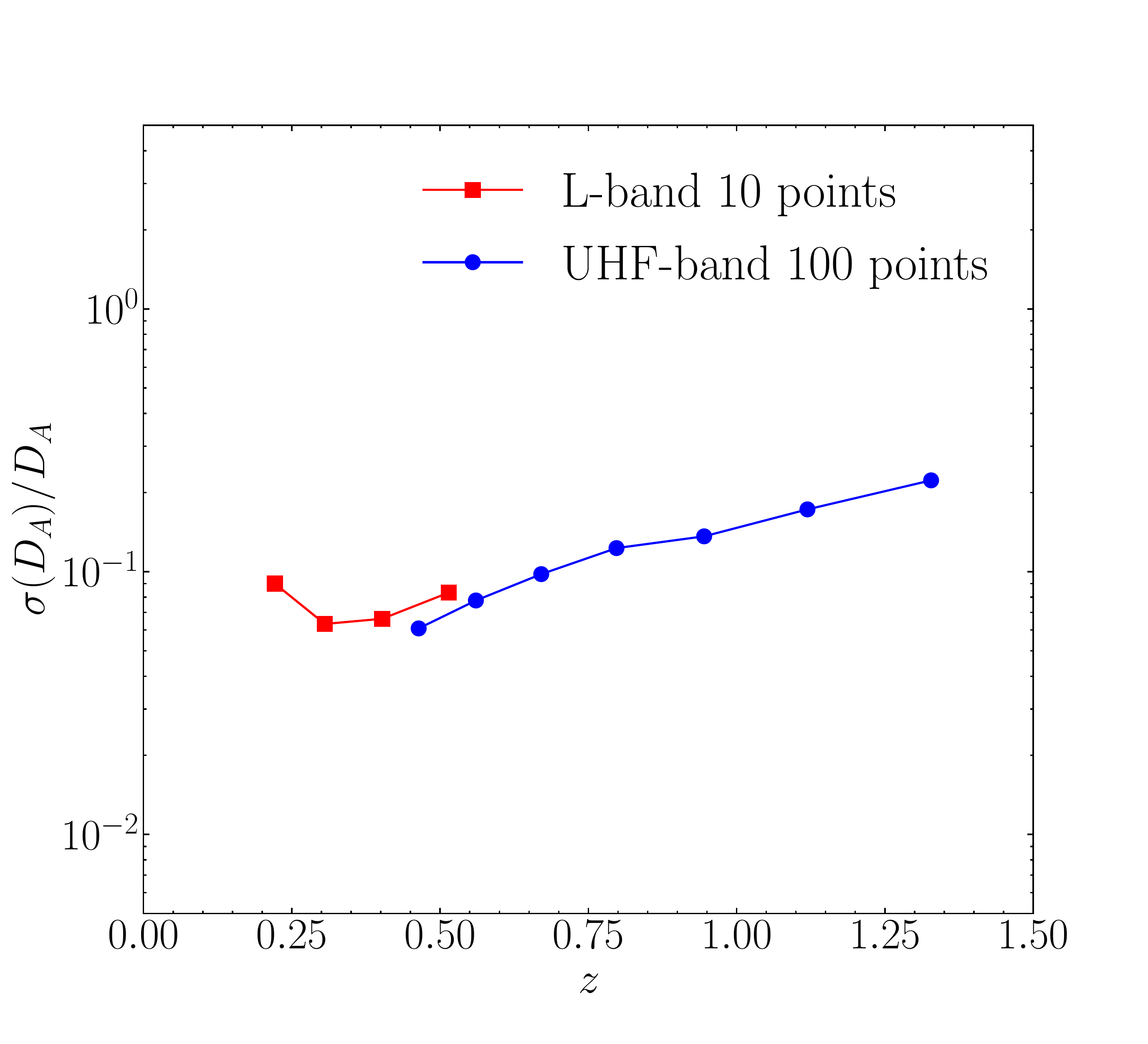}
  	\end{minipage}
  	\begin{minipage}{0.3\linewidth}
	\centering
  	    \includegraphics[width=\textwidth]{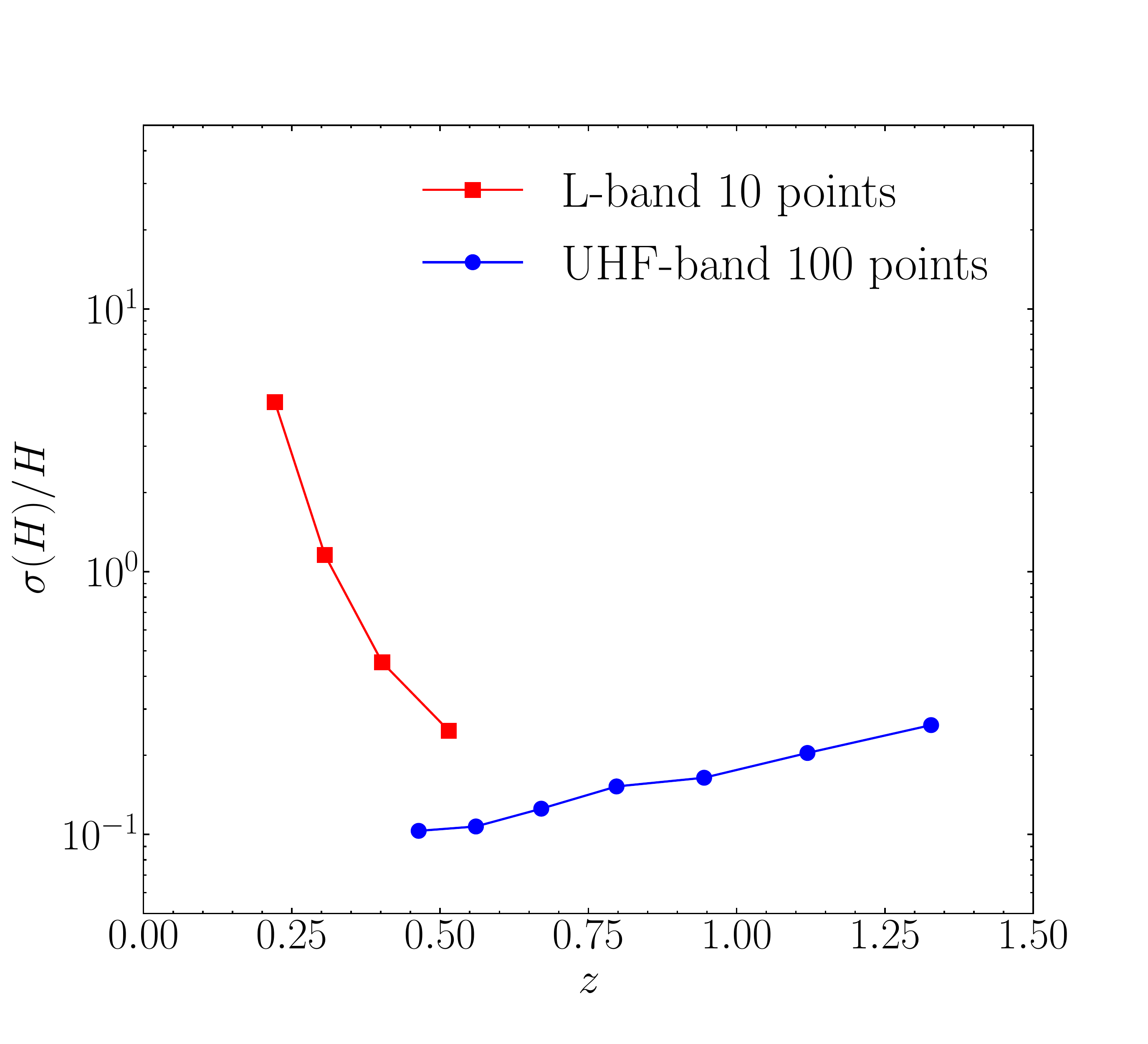}
  	\end{minipage}
  	\begin{minipage}{0.3\linewidth}
	\centering
  	    \includegraphics[width=\textwidth]{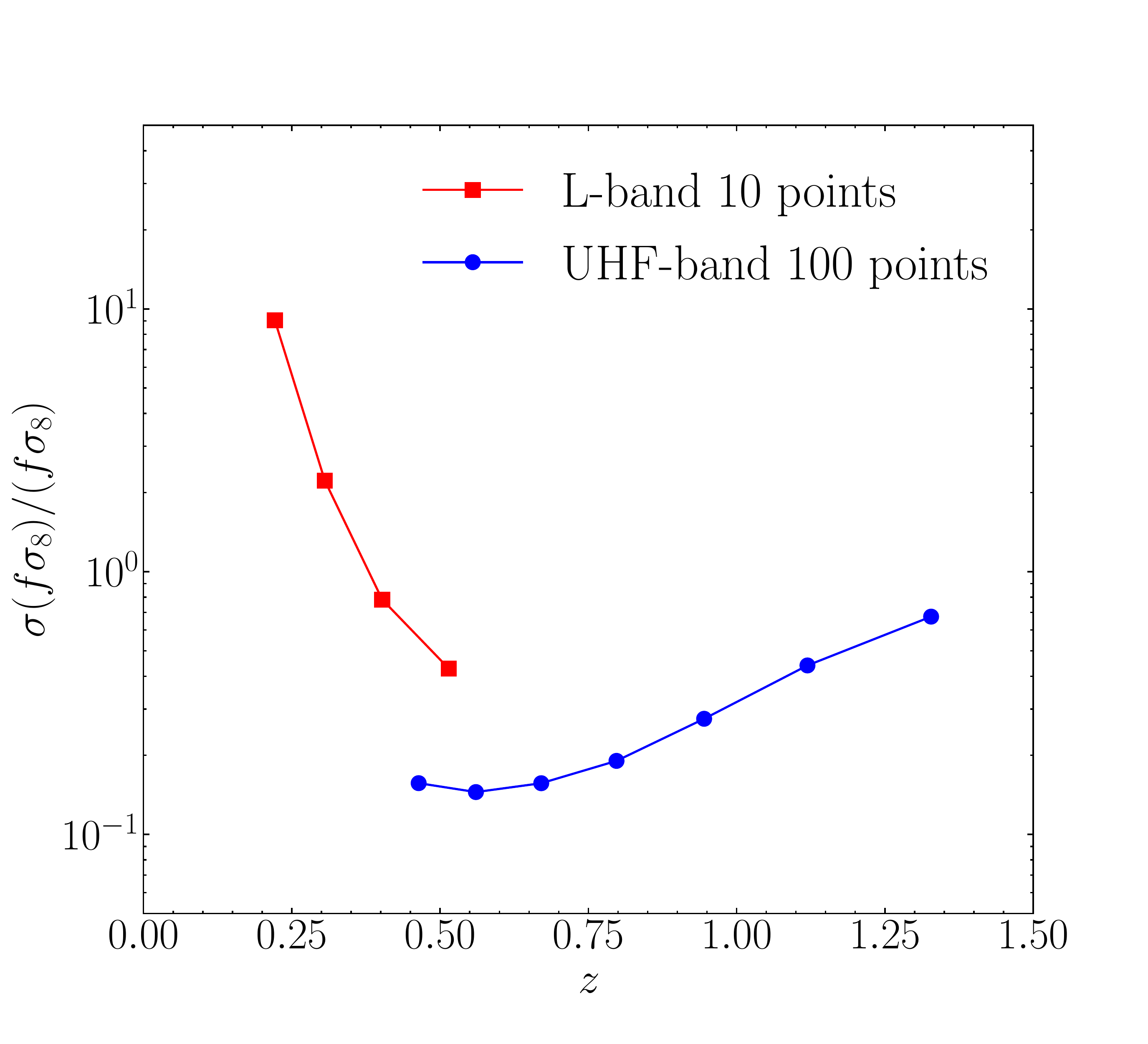}
  	\end{minipage}
    \caption{Fractional errors on $D_A(z)$, $H(z)$ and $f\sigma_8(z)$ obtained with MeerKAT L-band and UHF-band.}
    \label{fig:bao_error}
\end{figure*}

In this subsection, we explore the capability of MeerKAT \hi\ IM survey with interferometer mode of constraining cosmological parameters. 

Given the power spectrum measurement at a given redshift, the Fisher matrix for a set of observables $\{p_i\}$ can be written as
\begin{equation}
F_{i j} = \frac{1}{8\pi^2} V_{\rm{bin}} \int_{-1}^{1} {\rm d}\mu \int_{k_{\rm{min}}}^{k_{\rm{max}}} k^2 {\rm d} k \frac{\partial P_{\rm{tot}}}{\partial p_i} \frac{\partial P_{\rm{tot}}}{\partial p_j}. \label{fisher}
\end{equation}
Here, we take the set of observables $\{p_i(z_n)\}$ as \{$D_A(z_n)$, $H(z_n)$,  $f\sigma_8(z_n)$, $b\sigma_8(z_n)$, $\sigma_{\rm{NL}}$\} in each redshift bin $z_n$. The nuisance parameters $b\sigma_8(z_n)$ and $\sigma_{\rm{NL}}$ can be marginalized by selecting the submatrix of $F_{i j}^{-1}$ with only the appropriate columns and rows. Therefore, we can derive the measurement errors on $D_A(z)$, $H(z)$ and $f\sigma_8(z)$. 

For MeerKAT L-band (900-1200 MHz) and UHF-band (580-1000 MHz), we divide these frequency bands into some bins with equal bandwidth $\Delta \nu = 60 ~\rm{MHz}$ and then obtain the estimates for the measurement errors on observables in the corresponding redshift bins. We plot the fractional measurement errors on $D_A(z)$, $H(z)$ and $f\sigma_8(z)$ with MeerKAT 10000 h observation in Fig.~\ref{fig:bao_error}. We find that the survey with interferometer mode has a better measurement on $D_A(z)$, of which the fractional errors can reach roughly $10\%$ for MeerKAT L-band and UHF-band. Comparatively speaking, the fractional measurement errors on $H(z)$ and $f\sigma_8(z)$ seem slightly larger, though MeerKAT UHF-band performs slightly better than MeerKAT L-band.

In each redshift bin, the quantities $D_A(z)$, $H(z)$ and $f\sigma_8(z)$ are in some correlation, and thus we should take the correlation into account in the cosmological parameter inference. In a redshift bin, the $\chi^2$ function is given by 
\begin{align}
\chi^2 = \sum_{ij}{x}_i{F}_{ij}{x}_j,
\end{align}
where the Fisher matrix $\boldsymbol{F}$ serves as the inverse covariance matrix concerning $\{D_{\rm A}(z)$, $H(z)$, $[f\sigma_8](z)\}$, and $\boldsymbol{x}=(D_{{\rm A}}^{\rm th}(\boldsymbol{\xi})-D_{{\rm A}}^{\rm obs},~H^{\rm th}(\boldsymbol{\xi})-H^{\rm obs},~{[f\sigma_8]}^{\rm th}(\boldsymbol{\xi})-{[f\sigma_8]}^{\rm obs})$, with $\boldsymbol{\xi}$ denoting a set of cosmological parameters (such as \{$\Omega_{\rm m}$, $H_0$, $w/w_0$, $w_a$\}). Here, the superscripts ``th'' and ``obs'' represent the theoretical values in a cosmological model and the mock observational data in our simulation, respectively. The total $\chi_{\rm tot}^2$ function is the sum of $\chi^2$ of all redshift bins. Furthermore, we adopt the Markov Chain Monte Carlo (MCMC) approach to maximize the likelihood $\mathcal{L}\propto \exp{(-\chi^2/2)}$ to infer the posterior probability distributions of the cosmological parameters. We use the publicly accessible \texttt{CosmoMC} code to perform the MCMC analysis. In the cosmological parameter inference, we also employ the CMB angular power spectra data of \emph{Planck} 2018 TT,TE,EE+lowE.

%with respect to the cosmological parameters (such as \{$\Omega_{\rm m}$, $H_0$, $w/w_0$, $w_a$\}), 

\begin{figure}
	\centering
  	\begin{minipage}{0.95\linewidth}
	\centering
  	    \includegraphics[width=\textwidth]{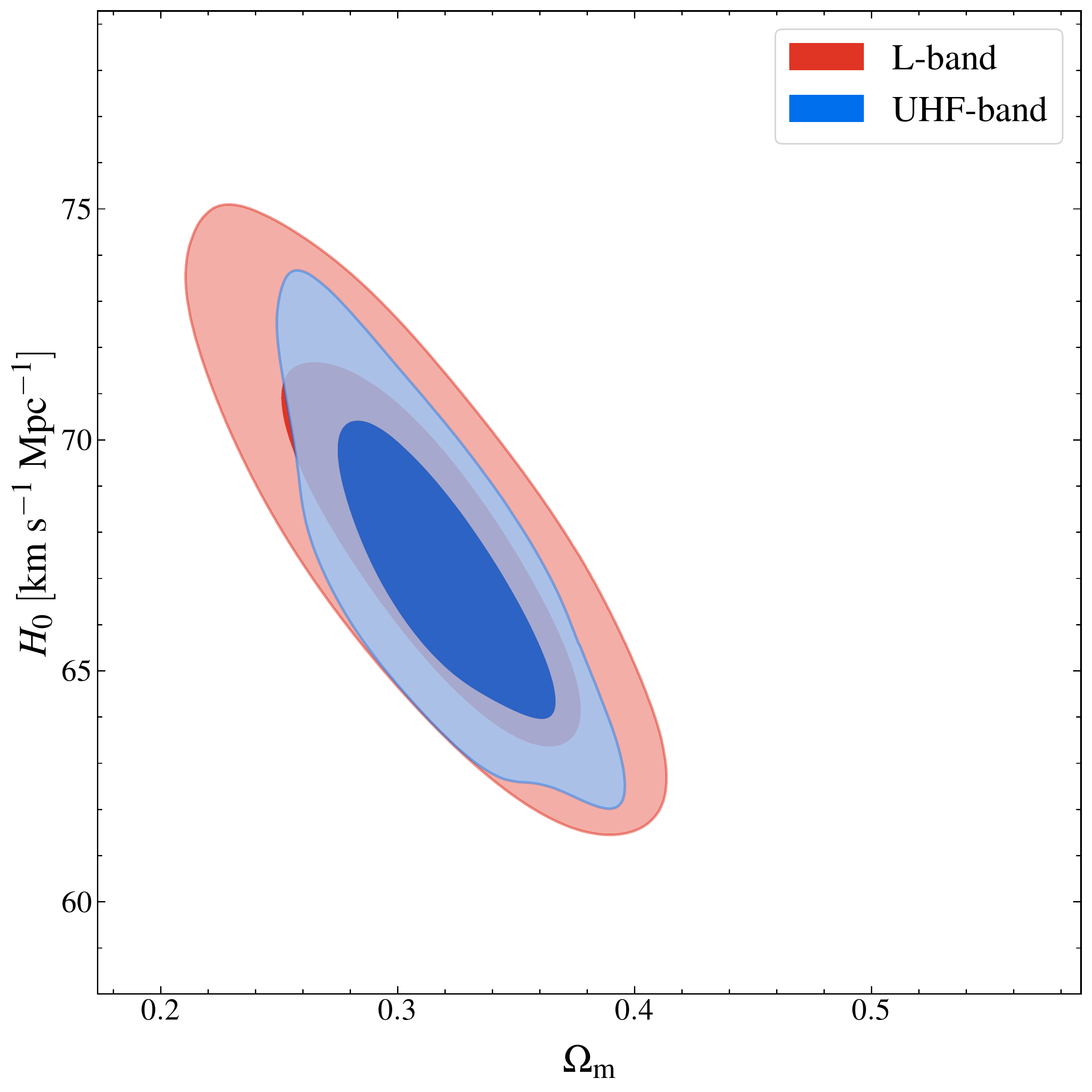}
  	\end{minipage}
    \caption{Constraints on $\Omega_{\rm m}$ and $H_0$ with MeerKAT L-band and UHF-band in the $\Lambda$CDM model.}
    \label{fig:LCDM}
\end{figure}

\begin{figure}
	\centering
  	\begin{minipage}{0.95\linewidth}
	\centering
  	    \includegraphics[width=\textwidth]{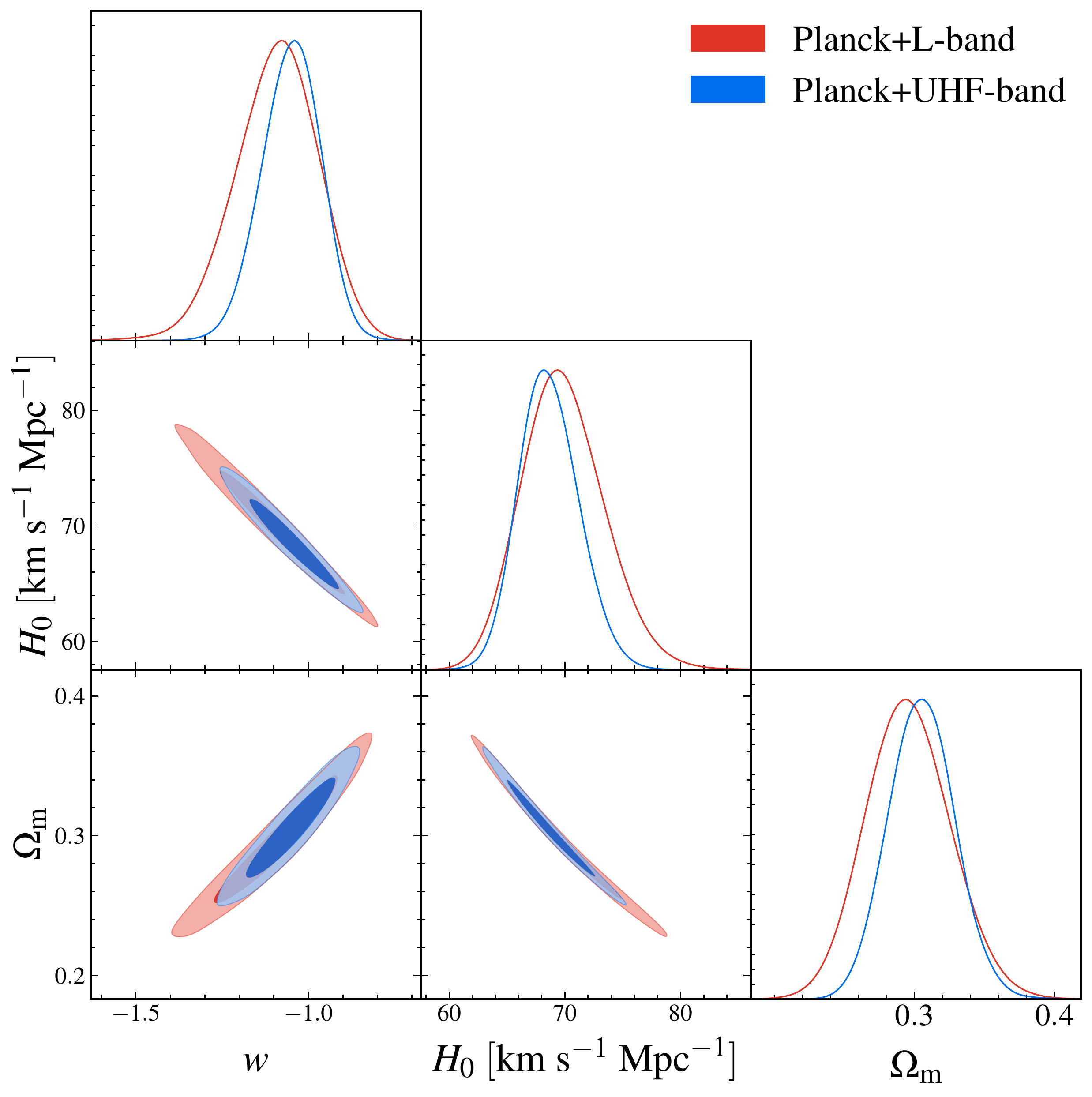}
  	\end{minipage}
    \caption{Constraints on $\Omega_{\rm m}$, $H_0$ and $w$ with MeerKAT L-band and UHF-band in combination with \emph{Planck} data in the $w$CDM model.}
    \label{fig:wCDM}
\end{figure}

\begin{figure}
	\centering
  	\begin{minipage}{0.95\linewidth}
	\centering
  	    \includegraphics[width=\textwidth]{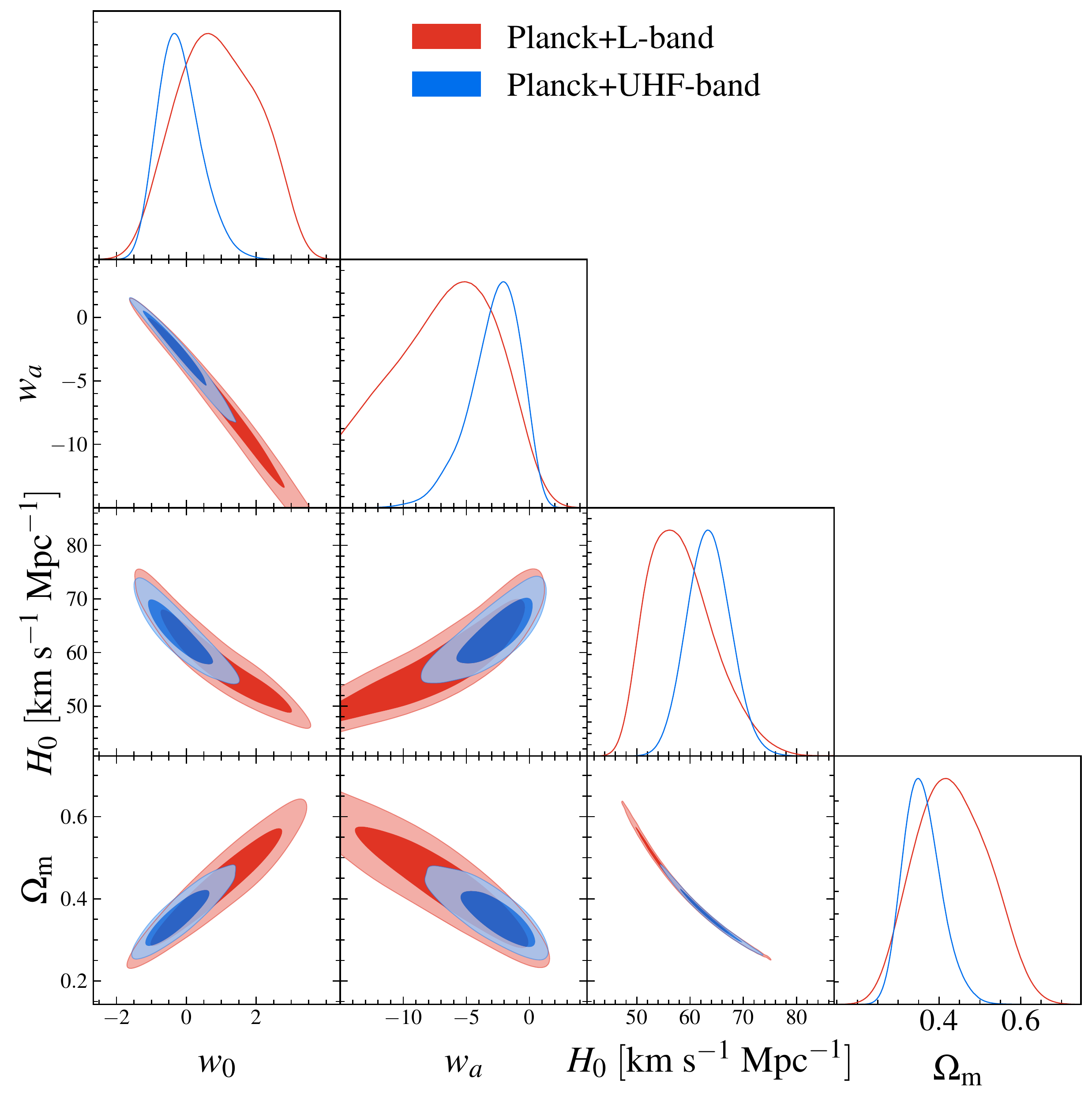}
  	\end{minipage}
    \caption{Constraints on $\Omega_{\rm m}$, $H_0$, $w_0$ and $w_a$ with MeerKAT L-band and UHF-band in combination with \emph{Planck} data in the $w_0w_a$CDM model.}
    \label{fig:CPL}
\end{figure}

%\blue{To place quantitative constraints on the cosmological parameters, we use the Fisher information matrix for $\{D_{\rm A}(z_n)$, $H(z_n)$, $[f\sigma_8](z_n); n=1,..., N\}$ in $N$ redshift bins. %The covariance matrix of the parameters is given by the inverse of the Fisher matrix. And then we adopt the Markov Chain Monte Carlo (MCMC) approach to maximize the likelihood $\mathcal{L}\propto \exp{(-\chi^2/2)}$ to infer the posterior probability distributions of the cosmological parameters. The $\chi^2$ function for a redshift bin can be defined as
%\begin{align}
%\chi^2 = \sum_{ij}{x}_i{F}^n_{ij}{x}_j,
%\end{align}
%where $\boldsymbol{x}=(H^{\rm th}(z)-H^{\rm obs},\, D_{{\rm A}}^{\rm th}(z)-D_{{\rm A}}^{\rm obs},\, {[f\sigma_8]}^{\rm th}(z)-{[f\sigma_8]}^{\rm obs})$. The total $\chi^2$ function is the sum of the $\chi^2$ of all redshift bins. Then, we insert it into the publicly accessible \texttt{CosmoMC} code to perform the MCMC analysis. Meantime, we also employ the CMB angular power spectra data of \emph{Planck} 2018 TT,TE,EE+lowE in the analysis.}
%Next, from the cosmological measurements on $D_A(z)$, $H(z)$ and $f\sigma_8(z)$, we can constrain the various dark enenrgy models, including the $\Lambda$CDM, $w$CDM and $w_0w_a$CDM models, by performing a Markov Chain Monte Carlo (MCMC) analysis. 
The $1 \sigma$ errors of the cosmological parameters are summarized in Table~\ref{tab1}. The $1 \sigma$ and $2 \sigma$ posterior distribution contours of cosmological parameters are shown in Figs.~\ref{fig:LCDM}--\ref{fig:CPL}.

%Next, from the cosmological measurement on $D_A(z)$, $H(z)$ and $f\sigma_8(z)$, we can constrain the Hubble constant $H_0$ and DE by performing a Markov Chain Monte Carlo (MCMC) analysis. The constraint results on $H_0$ in the flat $\Lambda$CDM model are showed in Fig.~\ref{fig:LCDM}. We obtain $H_0 = 66.6\pm4.5 $ km s$^{-1}$ Mpc$^{-1}$ with MeerKAT L-band and $H_0 = 66.4^{+3.2}_{-2.6}$ km s$^{-1}$ Mpc$^{-1}$ with MeerKAT UHF-band. %It is found that MeerKAT UHF-band performs better than L-band in constraining $H_0$. MeerKAT UHF-band, giving $1\sigma$ error $\sigma(H_0)=2.9$ km s$^{-1}$ Mpc$^{-1}$, is comparable to BINGO that refers to the analysis from \citet{Wu:2021vfz}.
%It is found that MeerKAT UHF-band performs better than L-band in constraining $H_0$ with giving $1\sigma$ error $\sigma(H_0)=2.9$ km s$^{-1}$ Mpc$^{-1}$, which is comparable to BINGO that refers to the analysis from \citet{Wu:2021vfz}. In addition, our result is better than that with MeerKAT UHF-band 4000 h observation of the single-dish mode \citep{Cunnington:2022ryj}. In contrast, also using interferometer mode, Tianlai and CHIME are more powerful of constraining $H_0$ than MeerKAT, because they have more compact short baselines \citep{Wu:2021vfz}.

\begin{table*}%[htbp]
\small
\setlength\tabcolsep{6pt}
\renewcommand{\arraystretch}{1.5}
\caption{\label{tab1} {The $1\sigma$ errors of the cosmological parameters in the $\Lambda$CDM, $w$CDM, and $w_0w_a$CDM models using \emph{Planck}, MeerKAT L-band and UHF-band. Note that here $H_0$ is in units of ${\rm km\ s^{-1}\ Mpc^{-1}}$.}}
\centering
\begin{tabular}{ccccccccccccccc}
\hline\multirow{2}{*}{Error} &\multicolumn{3}{c}{$\Lambda$CDM}& & \multicolumn{3}{c}{$w$CDM}& & \multicolumn{3}{c}{$w_0w_a$CDM}\\
 \cline{2-4}\cline{6-8}\cline{10-12}
 &\emph{Planck} &L-band &UHF-band&  &\emph{Planck} &\emph{Planck}+L-band &\emph{Planck}+UHF-band&  &\emph{Planck} &\emph{Planck}+L-band &\emph{Planck}+UHF-band& \\
\hline
$\sigma(\Omega_{\rm m})$
                   & $0.008$
                   & $0.044$
                   & $0.028$&
                   & $0.034$
		          &$0.030$
                   & $0.024$&
                   & $0.103$
                   & $0.092$
		          &$0.046$&
\\
$\sigma(H_0)$
                   & $0.59$
                   & $2.8$
                   & $2.0$&
                   & N/A
		           &$3.5$
                   & $2.6$&
                   & N/A
                   & $6.1$
		          &$4.1$&
\\
$\sigma(w)$
                    & $-$
                    & $-$
                    & $-$&
                    & $0.25$
		            &$0.12$
                    & $0.08$&
                    & $-$
                    & $-$
		          &$-$&
\\
$\sigma(w_0)$
                    & $-$
                    & $-$
                    & $-$&
		          &$-$
                    & $-$
                    & $-$&
                    & $1.2$
                    & $1.1$
		        &$0.6$&
\\
$\sigma(w_a)$
                   & $-$
                   & $-$
                   & $-$&
		          &$-$
                    & $-$
                   & $-$&
                   & N/A
                   & $4.3$
		          &$2.0$&
\\
\hline
\end{tabular}
\centering
\end{table*}

In the $\Lambda$CDM model, we obtain $\sigma(\Omega_{\rm m})=0.044$ and $\sigma(H_0)=2.8~{\rm km\ s^{-1}\ Mpc^{-1}}$ with MeerKAT L-band and $\sigma(\Omega_{\rm m})=0.028$ and $\sigma(H_0)=2.0~{\rm km\ s^{-1}\ Mpc^{-1}}$ with MeerKAT UHF-band. We find that UHF-band performs better than L-band in constraining $\Omega_{\rm m}$ and $H_0$. Recently, \citet{Cunnington:2022ryj} gave a result of $H_0 = 69.1^{+8.4}_{-5.7}~{\rm km\ s^{-1}\ Mpc^{-1}}$ from the position of the turnover location with MeerKAT UHF-band 4000 h survey with single-dish mode. It is found that a better constraint on $H_0$ with MeerKAT UHF-band interferometer mode can be given although a longer observational time of 10000 h is needed.
In comparison with other radio telescopes, MeerKAT L-band and BINGO perform similarly, while MeerKAT UHF-band performs nearly as well as FAST in constraining $\Omega_{\rm m}$ and $H_0$ in the $\Lambda$CDM model \citep{Wu:2021vfz}. Compared to the Stage-III dark energy experiments, such as DES, we find that MeerKAT UHF-band gives a smaller error on $\Omega_{\rm m}$ than DES with $\Omega_{\rm m}=0.339^{+0.032}_{-0.031}$ in the $\Lambda$CDM model \citep{DES:2020ahh}.

In the $w$CDM model, in order to help break the parameter degeneracy, we combine MeerKAT \hi IM survey with \emph{Planck} TT,TE,EE+lowE power spectrum \citep{Planck:2018vyg} in the MCMC analysis. The $1\sigma$ and $2\sigma$ measurement error contours for $\Omega_{\rm m}$, $H_0$ and $w$ are shown in Fig.~\ref{fig:wCDM}. We obtain $\sigma(\Omega_{\rm m})=0.030$, $\sigma(H_0)=3.5~{\rm km\ s^{-1}\ Mpc^{-1}}$ and $\sigma(w)=0.12$ with \emph{Planck}+L-band and $\sigma(\Omega_{\rm m})=0.024$, $\sigma(H_0)=2.6~{\rm km\ s^{-1}\ Mpc^{-1}}$ and $\sigma(w)=0.08$ with \emph{Planck}+UHF-band. It can be seen that MeerKAT UHF-band combined with \emph{Planck} data gives tighter constraints on cosmological parameters in the $w$CDM model, with the conclusion the same as in the $\Lambda$CDM model. We find that MeerKAT has a very limited capability of constraining $w$, and the error on $w$ is still larger even in combination with \emph{Planck} data. Given that current cosmological observations favor a value of $w = -1$, as indicated by the result of $w=-1.028\pm0.031$ obtained from \emph{Planck}+BAO+Supernova \citep{Planck:2018vyg}, it remains challenging to distinguish between $w$CDM and $\Lambda$CDM using MeerKAT \hi IM survey.

%\blue{Considering that current cosmological observations prefer $w = -1$, such as the result of $w=-1.028\pm0.031$ from Planck+BAO+Supernova \citep{Planck:2018vyg}, so we still can not distinguish between $w$CDM and $\Lambda$CDM using MeerKAT \hi IM survey.}

Finally, we forecast the constraints on cosmological parameters in the $w_0w_a$CDM model. The $1\sigma$ and $2\sigma$ measurement error contours are shown in Fig.~\ref{fig:CPL}. We focus on the dark-energy equation of state parameters $w_0$ and $w_a$. We obtain $\sigma(w_0)=1.1$ and $\sigma(w_a)=4.3$ with \emph{Planck}+L-band, and $\sigma(w_0)=0.6$ and $\sigma(w_a)=2.0$ with \emph{Planck}+UHF-band. 
%Note that dark energy dominates the evolution of the universe in the redshift range of $z\lesssim 0.4$. MeerKAT L-band has very limited constraining power for dark energy at the range of $0.18<z<0.58$, and MeerKAT UHF-band only surveys at the range of $0.42<z<1.45$. 
%It is evident that the MeerKAT \hi IM survey in interferometer mode does not yield tight constraints on the time-varying equation of state of dark energy. 
In comparison, the most advanced constraints on $w_0$ and $w_a$ are derived from the \emph{Planck}+BAO+Supernova data \citep{Planck:2018vyg}, providing an $8\%$ constraint on $w_0$ and an error of 0.3 for $w_a$. 
Therefore, it is not recommended to use interferometer mode \hi IM survey with MeerKAT
to study the cosmological evolution of the dark-energy equation of state.
On the other hand, by utilizing the single-dish mode, MeerKAT could potentially
achieve a large survey volume and more stringent constraints. 
Additionally, we can anticipate that future larger radio telescopes, such as SKA, will facilitate precise measurements of dark energy \citep{Wu:2021vfz,Wu:2023wpj,Wu:2022jkf}.

%The $1\sigma$ and $2\sigma$ measurement error contours for DE equation of state parameters $w_0$ and $w_a$ are showed in Fig.~\ref{fig:CPL}. We find that only Planck data can not constrain $w_0$ and $w_a$ very well. When combining \hi\ IM survey, we obtain $\sigma(w_0)=1.4$ and $\sigma(w_a)=5.0$ with Planck+MeerKAT L-band and $\sigma(w_0)=1.3$ and $\sigma(w_a)=3.7$ with Planck+MeerKAT UHF-band, which is larger errors than that with BINGO \citep{Wu:2021vfz}. The period of DE dominance of the universe is around the redshift $z<0.4$.  MeerKAT L-band has a very limited constraint at the range of $0.18<z<0.58$ and MeerKAT UHF-band only surveys at the range of $0.42<z<1.45$. Therefore, MeerKAT can not give a strict constraint on DE. But we should keep optimistic that is because the precise measurement of DE can be achieved by the future larger radio telescopes, such as Tianlai, HIRAX and SKA \citep{Xu:2014bya,Wu:2021vfz,2022arXiv220209726W}.

\section{Conclusions}\label{sec:conclusion}

In this work, we give a detailed analysis on measuring the \hi\ IM delay power spectrum using the MeerKAT interferometer mode. We also discuss the capability of MeerKAT interferometer mode of constraining cosmological parameters. 

%with MeerKAT \hi\ IM survey. MeerKAT interferometer mode are employed by using delay spectrum which can avoid the foregrounds contamination. Then we assess the capability of constraining on the power spectrum and cosmological parameters with MeerKAT L-band and UHF-band respectively.

We use the Fisher matrix method to estimate the \hi\ power spectrum with MeerKAT IM observation. We find that the different survey fields have distinct impacts on determining the power spectrum errors in the limited observational time of 10 hours. As the observational time increases from 10 h to 10000 h, the power spectrum errors are reduced evidently until the cosmic variance begins to dominate. We also discuss the different survey strategies and find that the lower fractional errors on power spectrum at low $k$ are obtained when tracking 100 points for L-band and tracking 10 points for UHF-band in a total 10000 h observation.
%We used the Fisher matrix method to estimate the power spectrum with MeerKAT \hi\ IM observation. We find that the power spectrum error are reduced distinctly as the integration time increasing from 10 h to 10000 h until cosmic variance dominating. In addition, the survey field has the distinctly impacts of the power spectrum error.

We obtain the measurement errors on $D_A(z)$, $H(z)$ and $f\sigma_8(z)$ by using the Fisher matrix, and then use these measurements to constrain cosmological parameters in typical dark energy models, including $\Lambda$CDM, $w$CDM and $w_0w_a$CDM models, by performing the MCMC analysis. We obtain $\sigma(\Omega_{\rm m}) = 0.028$ and $\sigma(H_0) = 2.0$ km s$^{-1}$ Mpc$^{-1}$ with MeerKAT UHF-band which are better than the results of $\sigma(\Omega_{\rm m}) = 0.044$ and $\sigma(H_0) = 2.8 $ km s$^{-1}$ Mpc$^{-1}$ with MeerKAT L-band in the $\Lambda$CDM model. However, MeerKAT has a very limited constraining power for the dark-energy equation of state, such as $w$ in the $w$CDM model and $w_0$ and $w_a$ in the $w_0w_a$CDM model, even though in combination with \emph{Planck} data. 
%Obtaining the measurement errors on $D_A(z)$, $H(z)$ and $f\sigma_8(z)$ by the Fisher matrix, we then use it to constrain $H_0$ and DE by performing the MCMC analysis. We get $\sigma(H_0) = 2.9$ km s$^{-1}$ Mpc$^{-1}$ with MeerKAT UHF-band which is better than the result of $\sigma(H_0) = 4.5 $ km s$^{-1}$ Mpc$^{-1}$ with MeerKAT L-band. Combining with Planck data, we achieve $\sigma(w_0)=1.4$ and $\sigma(w_a)=5.0$ with Planck+MeerKAT L-band and $\sigma(w_0)=1.3$ and $\sigma(w_a)=3.7$ with Planck+MeerKAT UHF-band.

Though MeerKAT L-band and UHF-band \hi\ IM surveys in interferometer mode have very limited constraining power for dark energy, our analysis still provides a useful guide for the near future MeerKAT survey. It is expected that future larger radio telescope arrays, such as SKA, will have a much better and more powerful performance in cosmological research. In addition, MeerKAT baselines are not short enough for detecting large cosmological scales, but the measurements with MeerKAT interferometer mode on these scales are still very useful in detecting \hi\ content of galaxies, obtaining the cross-correlation between \hi\ content and star formation rates \citep{Wolz:2015ckn}, constraining warm dark matter \citep{Carucci:2015bra} and breaking the degeneracy between $\Omega_{\hi}$ and $b_{\hi}$ \citep{Chen:2020uld}. These aspects deserve further detailed investigations in the future.

\section*{Acknowledgements}
We thank Peng-Ju Wu and Li-Yang Gao for the helpful discussions. 
This work was supported by the National SKA Program of China (Grants Nos.
2022SKA0110200 and 2022SKA0110203), the National Natural Science Foundation of China (Grants
Nos. 11975072, 11875102, and 11835009), and the 111 Project (Grant No. B16009).

%%%%%%%%%%%%%%%%%%%%%%%%%%%%%%%%%%%%%%%%%%%%%%%%%%
\section*{Data Availability}

 The data underlying this article will be shared on reasonable request to the corresponding author.

\bibliographystyle{mnras}
\bibliography{main} % if your bibtex file is called main.bib

% Alternatively you could enter them by hand, like this:
% This method is tedious and prone to error if you have lots of references
%\begin{thebibliography}{99}
%\bibitem[\protect\citeauthoryear{Author}{2012}]{Author2012}
%Author A.~N., 2013, Journal of Improbable Astronomy, 1, 1
%\bibitem[\protect\citeauthoryear{Others}{2013}]{Others2013}
%Others S., 2012, Journal of Interesting Stuff, 17, 198
%\end{thebibliography}

%%%%%%%%%%%%%%%%%%%%%%%%%%%%%%%%%%%%%%%%%%%%%%%%%%

%%%%%%%%%%%%%%%%%%%%%%%%%%%%%%%%%%%%%%%%%%%%%%%%%%

% Don't change these lines
\bsp	% typesetting comment
\label{lastpage}
\end{document}